\documentclass[prb, floatfix, superscriptaddress, longbibliography, reprint]{revtex4-2}
\usepackage[pdftex,plainpages=false,colorlinks=true,linkcolor=blue, citecolor=blue, urlcolor=blue]{hyperref}
\usepackage{amsfonts}
\usepackage{amsmath}
\usepackage{amssymb}
\usepackage{graphicx}% Include figure files
\usepackage{natbib}
\usepackage{color}
\usepackage{placeins}
\usepackage{physics}
\usepackage{orcidlink}
\usepackage{circledsteps}
\begin{document}

\title{Charge-transfer gap size and oxygen hole content as two mechanisms controlling $T_c$ in the Emery model}
\author{Eleanor M. O'Callaghan \orcidlink{0009-0006-0014-4153}}
\affiliation{Department of Physics, Royal Holloway, University of London, Egham, Surrey, UK, TW20 0EX}
\author{Nicolas Kowalski \orcidlink{0000-0002-7955-8394}}
\affiliation{D\'epartement de physique, Institut quantique \& RQMP, Universit\'e de Sherbrooke, Sherbrooke, Qu\'ebec, Canada J1K 2R1}
\author{A.-M. S. Tremblay \orcidlink{0000-0001-6932-8299}}
\affiliation{D\'epartement de physique, Institut quantique \& RQMP, Universit\'e de Sherbrooke, Sherbrooke, Qu\'ebec, Canada J1K 2R1}
\author{Giovanni Sordi \orcidlink{0000-0003-2481-7544}}
\email[corresponding author: ]{giovanni.sordi@rhul.ac.uk}
\affiliation{Department of Physics, Royal Holloway, University of London, Egham, Surrey, UK, TW20 0EX}
\date{\today}

\begin{abstract}
Investigating the drivers of superconducting critical temperature trends in cuprates is crucial for uncovering the mechanism of high-temperature superconductivity. Here we study this problem in the canonical model of the copper-oxygen plane, the Emery model, with cellular dynamical mean-field theory. Using the Zaanen-Sawatzky-Allen diagram as a guiding framework, we systematically quantify how the maximum superconducting critical temperature $T_c^{\rm max}$ depends on the copper-oxygen energy distance and on the local repulsion on the copper orbital. The latter is the dominant variable in varying $T_c^{\rm max}$. Unexpectedly, $T_c^{\rm max}$ is optimized not only near the charge-transfer insulator to metal boundary, consistent with previous findings, but also deep in the charge-transfer regime, revealing an unexplored mechanism. Then we link model parameters to physical observables, identifying the charge-transfer gap size and the oxygen hole content as two mechanisms controlling $T_c^{\rm max}$. $T_c^{\rm max}$ increases monotonically as the oxygen hole content increases and the charge gap size decreases. Our work provides predictions for proposed realizations of the Emery model with ultracold atoms and a theoretical framework for understanding key experimental trends in hole-doped cuprates. 
\end{abstract}
 
\maketitle

\section{Introduction} 
\label{sec:Introduction}
 
Understanding superconducting critical temperature trends in cuprates is vital to elucidating the mechanism of high-temperature superconductivity~\cite{Norman2011, keimerRev, ift}. 
Over the past decades, intense experimental efforts have identified systematic trends relating the superconducting transition temperature $T_c$ to a number of properties. These include superconducting properties (e.g. superfluid density~\cite{Uemura:PRL1989}) and magnetic properties (e.g. superexchange interaction~\cite{Wang:NatComm2022, Ruan:SciBull2016}), as well as crystal structure parameters (e.g. apical atoms~\cite{Peng:NatPhys2017, Davis:PNAS2022}, number of copper-oxygen layers~\cite{Wang:Science2023}), electronic structure parameters (e.g. charge gap size~\cite{Ruan:SciBull2016, Davis:PNAS2022, Wang:Science2023}), and other physical quantities (e.g. the occupation of copper and oxygen orbitals~\cite{Rybicki:NatComm2016, Jurkutat:PNAS2023}). While these trends show the complexity of controlling $T_c$ in cuprates, they also provide constraints and insights for understanding the key mechanisms of high-temperature superconductivity.

At theory level, over the years extensive studies based on first-principle methods and model calculations~\cite{Norman2011, Kotliar:PhysC1988, Kotliar:PRB1988, Ohta:PRB1991, Feiner:PRB1992, Scalapino:1994, Raimondi:PRB1996, Pavarini:PRL2001, Kent:PRB2008, ArrigoniCuO2, Weber2011, Lorenzo3band, Dash:PRB2019, Nicolas:PNAS2021, Mai:PRB2021, Mai:npj2021, Vucicevic:PRB2024, Cui:NatComm2025, BacqLabreuil:PRX2025, St-Cyr:2025, Vadnais:2026, Jacob:2026} have led to progress in linking empirical trends in $T_c$ to microscopic parameters. Despite advances, the mechanisms underlying these trends are still under scrutiny. In this context, the Emery model~\cite{Emery_1987, Varma_1987} provides a simple conceptual framework for revealing universal patterns in $T_c$ hidden inside the complexity of cuprate materials.

\begin{figure}[t!]
\centering{
\includegraphics[width=1.0\linewidth]{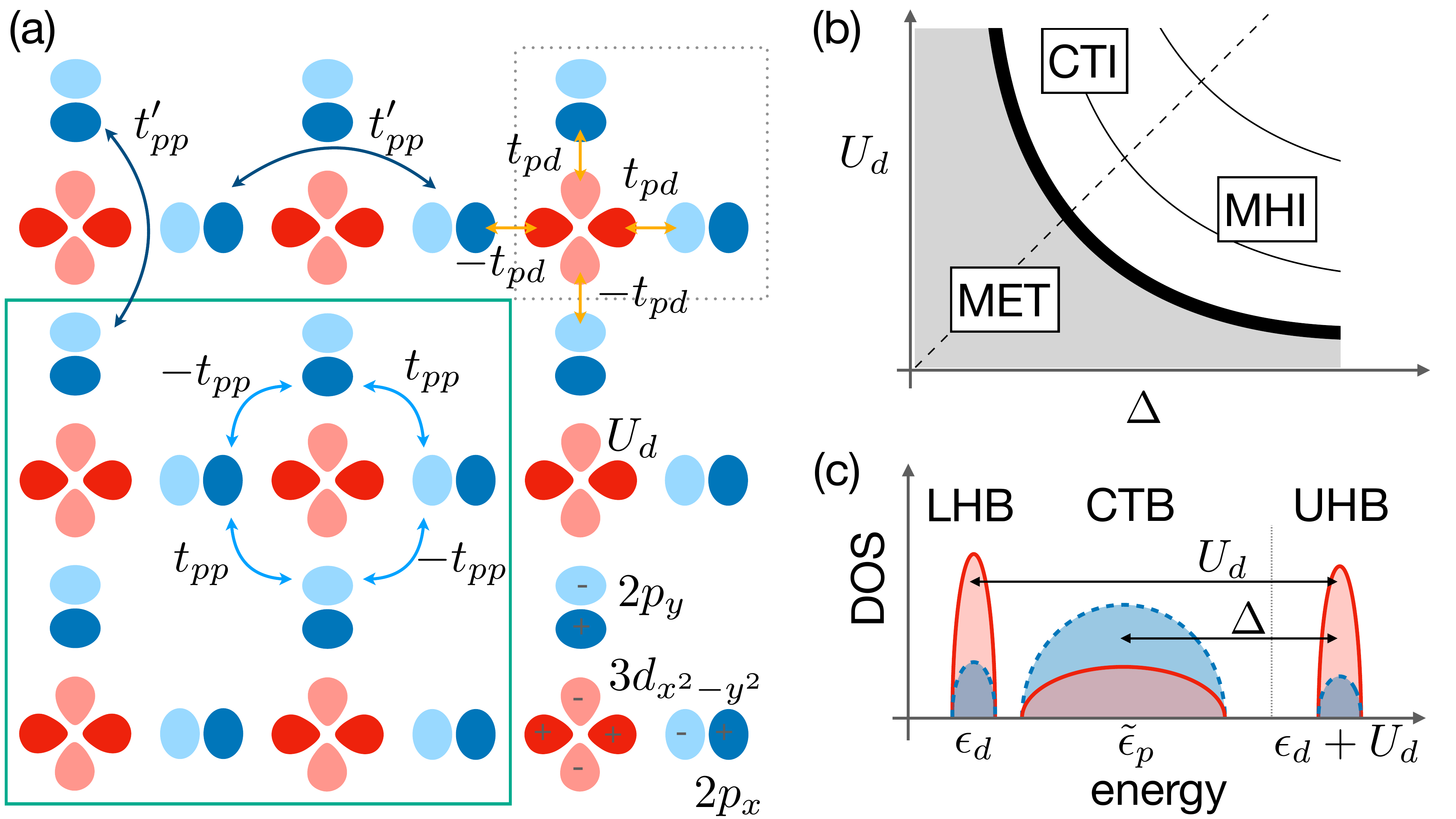}
}
\caption{(a) Sketch of the Emery model. The dotted gray square indicates the lattice unit cell, which consists of a Cu $3d_{x^2-y^2}$ orbital (red) of onsite energy $\epsilon_d$ and the O $2p_x$, $2p_y$ orbitals (blue) of onsite energy $\epsilon_p$. We use bright and pale colors to denote the positive and negative signs of the orbital lobes. We use the convention that the overlap between orbital lobes of opposite (equal) sign yields to a positive (negative) hopping amplitude. There are three hopping amplitudes: $t_{pd}$ (between the nearest-neighbor Cu-O orbitals) [orange arrows], $t_{pp}$ and $t_{pp}^\prime$ (between nearest-neighbor and next-nearest-neighbor O $2p_x$ and $2p_y$ orbitals, respectively) [light blue and dark blue arrows, respectively]. $U_d$ is the onsite Coulomb repulsion on the Cu orbitals. The solid green square indicates the $12$ site lattice cluster we use in our CDMFT calculations. 
(b) Sketch of the Zaanen-Sawatzky-Allen diagram, where $U_d$ is on the $y$ axis and the bare charge-transfer energy $\Delta = \epsilon_d + U_d -\tilde{\epsilon}_p$ is on the $x$ axis, where we renormalized $\epsilon_p$ by $\tilde{\epsilon}_p = \epsilon_p - 2t_{pp}$. $\Delta$ is the energy distance between the upper Hubbard band at $ \epsilon_d + U_d$ and the onsite energy of the O $2p$ orbitals, in the $t_{pd}\rightarrow 0$ limit. The thick black line seperates the metallic state (MET) from the correlated insulators (of Mott Hubbard (MHI) or charge-transfer (CTI) type). The thin black lines represent lines of constant charge gap size $\Delta_{\rm I}$.  
(c) Sketch of the Cu $3d$ (solid red line) and O $2p$ (dashed blue line) partial density of states of a charge-transfer insulator, for which $U_d > \Delta$. The Fermi level (dotted line) lies between the charge-transfer band (CTB) and the upper Hubbard band (UHB).
}
\label{fig:sketch}
\end{figure}

The Emery model is the canonical model of the electron behavior in the copper-oxygen plane in cuprates (see sketch in Fig.~\ref{fig:sketch}(a)). In its simplest form, it includes three orbitals per unit cell: two oxygen $2p_x$ and $2p_y$ orbitals that hybridize with one copper $3d_{x^2-y^2}$ orbital which has a strong local electron-electron repulsion. The Emery model has a large dimensional space defined by the hopping parameters between oxygen orbitals and between oxygen and copper orbitals, the onsite energies of the orbitals, and the strength of the Coulomb repulsion on the copper orbital. Therefore identifying the model parameters that optimize $T_c$ in such a large dimensional parameter space is a challenging task. This challenge is amplified by the strong electronic correlations on the copper orbital that require the use of nonperturbative methods~\cite{AMJulich}. 

A comprehensive exploration of the entire parameter space of the Emery model is still missing and nevertheless is key (i) for finding the loci of optimum $T_c$, (ii) for uncovering complex relations between model parameters, and (iii) for quantifying the relative importance of different model parameters in varying $T_c$. Our work is a step in this direction. 

To make progress, first we need a framework, or a map, that makes the Emery model multidimensional parameter space interpretable and enables an intuitive understanding of how the parameters of the Emery model relate. The Zaanen-Sawatzky-Allen diagram~\cite{zsa} provides such a framework, as we shall discuss in detail in this work. It maps the correlated states that can be obtained for five electrons (or, equivalently, one hole) per CuO$_2$ unit cell, with the oxygen orbitals fully filled and the copper orbital half-filled, in the extreme case of no copper-oxygen hybridization. As sketched in Fig.~\ref{fig:sketch}(b), it shows the onsite repulsion $U_d$ against the bare charge-transfer energy $\Delta$ (i.e. the energy distance between the upper Hubbard band and the onsite energy of the O orbitals, in the extreme case of no copper-oxygen hybridization). The relative size of $U_d$ and $\Delta$ sets the Emery model in an insulating state of Mott-Hubbard or charge-transfer type, or in a metallic state, if the kinetic energy gain overcomes the energy cost to excite an electron across the insulating gap. 

Having a map, to make further progress, we build on previous research. Specifically we focus on three main prior findings. First,  high-$T_c$ cuprates are doped charge-transfer insulators~\cite{Emery_1987, ift, keimerRev} (sketch of the density of states in Fig.~\ref{fig:sketch}(c)). Second, previous work places the parent state of hole-doped cuprates close to the charge-transfer insulator to metal boundary of the Zaanen-Sawatzky-Allen diagram, on the charge-transfer insulating side~\cite{Kotliar:IJMPB1991, Baumgartel:PRB1993, Scalapino:1994, Weber2011, Nicolas:PNAS2021}. Third, $T_c^{\rm max}$ decreases with increasing the charge-transfer gap size (contour lines of given charge gap size in Fig.~\ref{fig:sketch}(b)), i.e. moving away from the charge-transfer insulator to metal boundary of the Zaanen-Sawatzky-Allen diagram, as shown both experimentally~\cite{Ruan:SciBull2016, Davis:PNAS2022, Wang:Science2023} and theoretically~\cite{Weber2011, Nicolas:PNAS2021}. Overall, these insights allow us to confine our investigation to the region $U_d > \Delta$ of the Zaanen-Sawatzky-Allen diagram, close to the metal-insulator boundary, and to restrict our study to hole doping.

As a result, the idea of this work is to explore the large dimensional parameter space of the Emery model by fixing the hopping parameters and studying the variation of $T_c$ with three parameters: $U_d$, $\Delta$, and hole doping. Since for given $U_d$ and $\Delta$, the superconducting transition temperature $T_c$ forms a dome upon doping, we can map the maximum value of the superconducting dome, $T_c^{\rm max}$, as a function of just two model parameters: $U_d$ and $\Delta$. In this simpler two-dimensional parameter space, we find the loci of optimum $T_c^{\rm max}$. The second step of our study is then to connect the model parameters $U_d$ and $\Delta$ to physical observables, specifically the charge gap size and the density of holes in the oxygen orbital. In the parameter space spanned by these two physical observables, (i) we find the loci of optimum $T_c^{\rm max}$ and (ii) we uncover the dependencies between physical observables and $T_c^{\rm max}$. This allows us to compare model predictions to experimental trends~\cite{Rybicki:NatComm2016, Jurkutat:PNAS2023, Ruan:SciBull2016, Davis:PNAS2022, Wang:Science2023}.

This work  builds on our companion article~\cite{Eleanor_short}, where we show that for a given value of the charge gap size (i.e. along a contour line of given charge gap size in Fig.~\ref{fig:sketch}(b)), the optimal $T_c$ occurs for small charge-transfer energy that puts the model deep in the charge-transfer regime, characterized by $U_d \gg \Delta$. Put another way, in the companion paper~\cite{Eleanor_short} we solve a one-dimensional optimization problem, i.e. we maximize $T_c$ as a function of $\Delta$, for a fixed value of charge gap size $\Delta_{\rm I}$. Here we expand the companion article~\cite{Eleanor_short} by providing further detail of this result, by confirming that this result holds for another value of $\Delta_{\rm I}$, and most importantly by solving a two-dimensional optimization problem, where we maximize $T_c$ as a function of $\Delta$ and $U_d$. In terms of physical observables, this means that here we address the impact of the interplay between the charge gap size and oxygen hole content on $T_c$. 

This work is organised as follows. In Sec.~\ref{sec:Method} we describe the model and method. Sec.~\ref{sec:CTI} studies the parent charge-transfer insulator states of given charge gap sizes. Sec.~\ref{sec:SC} studies the superconducting state emerging from such insulating states. Sec.~\ref{sec:trends} examines the trends of the maximum superconducting temperature as a function of model parameters and physical observables, and contrasts model findings with experimental trends. Finally, Sec.~\ref{sec:summary} summarizes our results.  

\section{Model and Method}
\label{sec:Method}

The Emery model~\cite{Emery_1987} studied in this work is given by the Hamiltonian
\begin{align}
H & = \sum_{\mathbf{k} \sigma} C_{\mathbf{k} \sigma}^{\dagger}  \left[ \mathbf{h}_{0}(\mathbf{k}) -\mu \mathbf{I} \right]  C_{\mathbf{k} \sigma} 
+ U_d \sum_{\mathbf{R}_{i}} n_{d \mathbf{R}_{i} \uparrow} n_{d \mathbf{R}_{i} \downarrow} . 
\label{eq:EmeryModel}
\end{align}
Here, $C_{\mathbf{k} \sigma}^{\dagger} = ( d_{\mathbf{k} \sigma}^{\dagger} ,  p_{x \mathbf{k}  \sigma}^{\dagger} , p_{y \mathbf{k} \sigma}^{\dagger}  )$, $C_{\mathbf{k} \sigma} = \left( d_{\mathbf{k} \sigma},  p_{x \mathbf{k} \sigma}, p_{y \mathbf{k} \sigma} \right)^{T}$, where $d_{\mathbf{k} \sigma}^{(\dagger)}$ and $p_{\alpha \mathbf{k}  \sigma}^{(\dagger)}$, with $\alpha=x,y$, annihilate (create) an electron of wave vector $\mathbf{k}$ and spin $\sigma$ in the Cu $3d_{x^2-y^2}$ and O $2p_{\alpha}$ orbitals, respectively, and $n_{d \mathbf{R}_{i} \sigma} = d_{ \mathbf{R}_{i} \sigma}^{\dagger} d_{ \mathbf{R}_{i} \sigma}$ is the number operator for the Cu orbital at site $\mathbf{R}_{i}$. $U_d$ is the onsite Coulomb repulsion between electrons on the Cu orbital and $\mu$ is the chemical potential. The noninteracting part of the Hamiltonian is
\begin{align}
\mathbf{h}_0 (\mathbf{k}) & =
\left(
\begin{array}
[c]{ccc}
\epsilon_{d} & V_{dp_x} & V_{d p_y} \\
V^\dagger_{dp_x} & \tilde{\epsilon}_p + W_{p_x p_x} & W_{p_x p_y} \\
V^\dagger_{dp_y} & W^\dagger_{p_x p_y} & \tilde{\epsilon}_p + W_{p_y p_y} 
\end{array} \right)  .
\label{eq:h0}
\end{align}
Here, $V_{dp_\alpha} = t_{pd}\left(  1-e^{-ik_{\alpha}}\right)$, $W_{p_\alpha p_\alpha} = 2t_{pp}^\prime \cos k_{\alpha}$, and $W_{p_x p_y} = t_{pp}\left(  1-e^{ik_{x}}\right)  \left(  1-e^{-ik_{y}}\right)$, where we set the Cu-Cu lattice distance to unity. 
$t_{pd}$ is the hopping amplitude between nearest neighbor Cu-O orbitals, $t_{pp}$ and $t_{pp}^\prime$ are the hopping amplitudes between nearest neighbor and next nearest neighbor O-O orbitals. The phase factors of the hopping amplitudes are illustrated in Fig.~\ref{fig:sketch}(a). 
$\epsilon_{d}$ is the onsite energy of the Cu orbital, $\epsilon_{p} = \epsilon_{p_x} = \epsilon_{p_y}$ is the onsite energy of the O orbitals, which, following Ref.~\cite{AndersenLDA} we renormalize by $\tilde{\epsilon}_p = \epsilon_p -2t_{pp}$.  
 
We solve this model at finite temperature in the normal and $d$-wave superconducting state with the cellular extension~\cite{maier, kotliarRMP, tremblayR} of dynamical mean-field theory (CDMFT)~\cite{rmp}, following the procedure described in Ref.~\cite{Lorenzo3band}. CDMFT maps the Emery model onto a cluster impurity model embedded in a self-consistent reservoir of noninteracting electrons. Here we use the minimal cluster that describes $d$-wave superconductivity, i.e. a cluster of $12$ sites formed by $N_d=4$ Cu sites and $N_p=8$ O sites (green square in Fig.~\ref{fig:sketch}(a)). To solve the cluster quantum impurity problem we use the hybridization expansion continuous-time quantum Monte Carlo method (CT-HYB) in the segment representation, as detailed in Refs.~\cite{Lorenzo3band, Nicolas:PNAS2021, Nicolas:Master}. 

In the following, we set $t_{pp}=1$ as our unit of energy and temperature $T$ (with $k_B=1$). We study the model parameters $\epsilon_d=0$, $t_{pp}^\prime=1$, $t_{pd}=1.5$, as in Refs.~\cite{Lorenzo3band, Dash:PRB2019, Nicolas:PNAS2021, GiovanniPRB2025, GemmaPRB2025}, and we vary the Cu-O energy distance in the range $\tilde{\epsilon}_p -\epsilon_d \in [2, 10]$ and the interaction strength in the range $U_d \in [7, 16]$. These parameters set the Emery model in the charge-transfer regime. This can be understood as follows. For $t_{pd}=0$ and $U_d=0$, the Cu level lies below the degenerate O bands centered at $\tilde{\epsilon}_p$. A value of $U_d$ larger than $\tilde{\epsilon}_p -\epsilon_d$ splits the Cu level into two levels of energy $\epsilon_d$ and $\epsilon_d +U_d$. For a total occupation of $n_{\rm tot}= n_d +2n_p =5$, there is a charge gap between the degenerate O bands and the Cu level at $\epsilon_d +U_d$. A finite value of $t_{pd}$ turns the levels into bands and mixes the character of the Cu and O bands, resulting in a charge-transfer insulator, as sketched in Fig.~\ref{fig:sketch}(c).

\section{Charge-transfer insulators}
\label{sec:CTI}

This work examines the superconducting state that emerges when charge-transfer insulators are doped with holes. This section focuses on the parent charge-transfer insulating states. To classify these states, first we construct the Zaanen-Sawatzky-Allen diagram with the Emery model and then we characterize the charge-transfer insulating states of given charge gap sizes with the density of states. 

\subsection{Zaanen-Sawatzky-Allen diagram}

\begin{figure*}[t!]
\centering{
\includegraphics[width=1.0\linewidth]{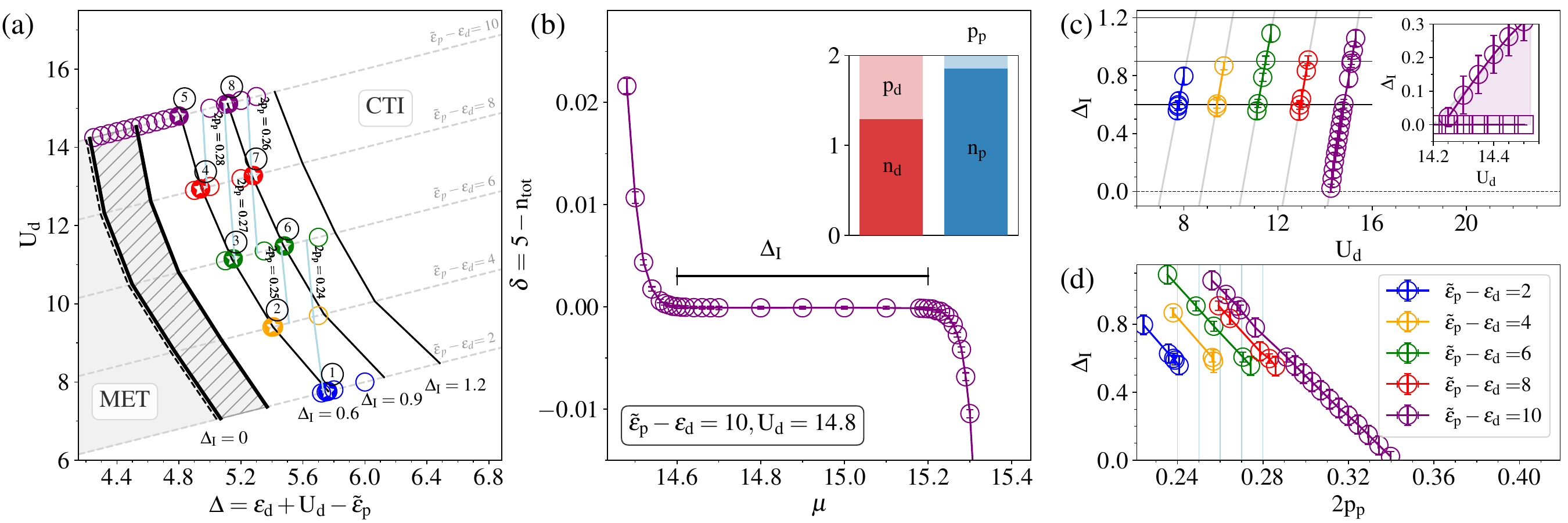}
}
\caption{(a) Zaanen-Sawatzky-Allen diagram $U_d$ vs $\Delta = \epsilon_d + U_d -\tilde{\epsilon}_p$ of the Emery model, computed with CDMFT.  Open circles indicate the parameters $(\Delta, U_d)$ investigated in our work. All data are at $n_{\rm tot}=5$, i.e. zero doping, and $T=1/50$. Gray dashed lines denote constant values of the bare Cu-O energy distance $\tilde{\epsilon}_p -\epsilon_d$. The data along each of these lines are assigned a different color. The hatched region bounded by thick black lines denotes the coexistence associated with the first-order metal-insulator transition. Thin black lines are lines of constant charge gap size $\Delta_{\rm I}$. The data that is closest to the lines of constant $\Delta_{\rm I}$ are marked by circles with a star, and labeled $1$ to $8$. The dashed black line marks where $\Delta_{\rm I}$ closes and is an estimate for the first-order transition. Pale blue lines are lines of constant oxygen hole content $2p_p$. 
(b) Doping $\delta=5-n_{\rm tot}$ vs chemical potential $\mu$ at $T=1/50$, for $(\tilde{\epsilon}_p -\epsilon_d=10, U_d=14.8)$ as a representative of our calculations. The plateau of $\delta(\mu)$ denotes the charge gap size $\Delta_{\rm I}$. Inset: partial occupancy of the orbitals at zero doping: $n_d\approx 1.291$, $p_d=2-n_d$ (dark and pale red, respectively); $n_p \approx 1.854$, $p_p=2-n_p$ (dark and pale blue, respectively). 
(c) Charge gap size $\Delta_{\rm I}$ vs interaction strength $U_d$ for different values of $\tilde{\epsilon}_p -\epsilon_d$. Intersection with horizontal lines at $\Delta_{\rm I}=0.6, 0.9, 1.2$ defines the lines of constant charge gap size in panel (a). Inset: magnified view of $\Delta_{\rm I}(U_d)$ for  $\tilde{\epsilon}_p -\epsilon_d=10$, showing hysteresis loop and the resulting coexistence region (shaded area) between a metallic (squares) and an insulating state (circles). 
(d) $\Delta_{\rm I}$ vs oxygen hole content $2p_p$. Intersection with vertical lines at selected values of $2p_p$ defines lines of constant O hole content in panel (a).
}
\label{fig:zsa}
\end{figure*}

Figure~\ref{fig:zsa}(a) shows the Zaanen-Sawatzky-Allen diagram of the Emery model computed with CDMFT at the low temperature $T=1/50$. This diagram has been calculated at the higher temperature $T=1/20$ in Ref.~\cite{GemmaPRB2025}. It describes the Emery model when the total number of electrons in the CuO$_2$ unit cell is equal to $5$. The Zaanen-Sawatzky-Allen diagram has the interaction strength on the Cu orbital $U_d$ on the $y$ axis and the bare charge-transfer energy $\Delta = \epsilon_d + U_d -\tilde{\epsilon}_p$ on the $x$ axis. $\Delta$ is the energy difference between the upper Hubbard band and the centre of the oxygen bands, in the atomic limit $t_{pd}\rightarrow 0$. Therefore, each point in the phase diagram is described by the coordinates $(\tilde{\epsilon}_p -\epsilon_d, U_d)$. 

Open circles show the model parameters investigated in this work. We chart out the phase diagram with scans along lines of constant bare Cu-O energy distance $\tilde{\epsilon}_p -\epsilon_d$ (gray diagonal lines). The assigned color code for the different values of $\tilde{\epsilon}_p -\epsilon_d$ is used consistently throughout this work. To map out the charge-transfer regime defined by $U_d > \Delta$, we study different values of $\tilde{\epsilon}_p -\epsilon_d$ ranging from $2$ to $10$. The Emery model would enter in the Mott-Hubbard regime where $U_d < \Delta$ upon further decreasing $\tilde{\epsilon}_p -\epsilon_d$ (see e.g. Ref.~\cite{GemmaPRB2025}).

The strategy of this work is to characterize each point of the $U_d-\Delta$ phase diagram with two physical quantities: (i) the charge-transfer gap size and (ii) the degree of electronic charge shared among copper and oxygen orbitals. This enables us to draw contour lines of these two quantities in Fig.~\ref{fig:zsa}(a) (thin black lines and pale blue lines, respectively), as we shall discuss below.

\subsubsection{Charge gap size $\Delta_{\rm I}$}

\begin{figure*}[t!]
\centering{\includegraphics[width=1.0\linewidth]{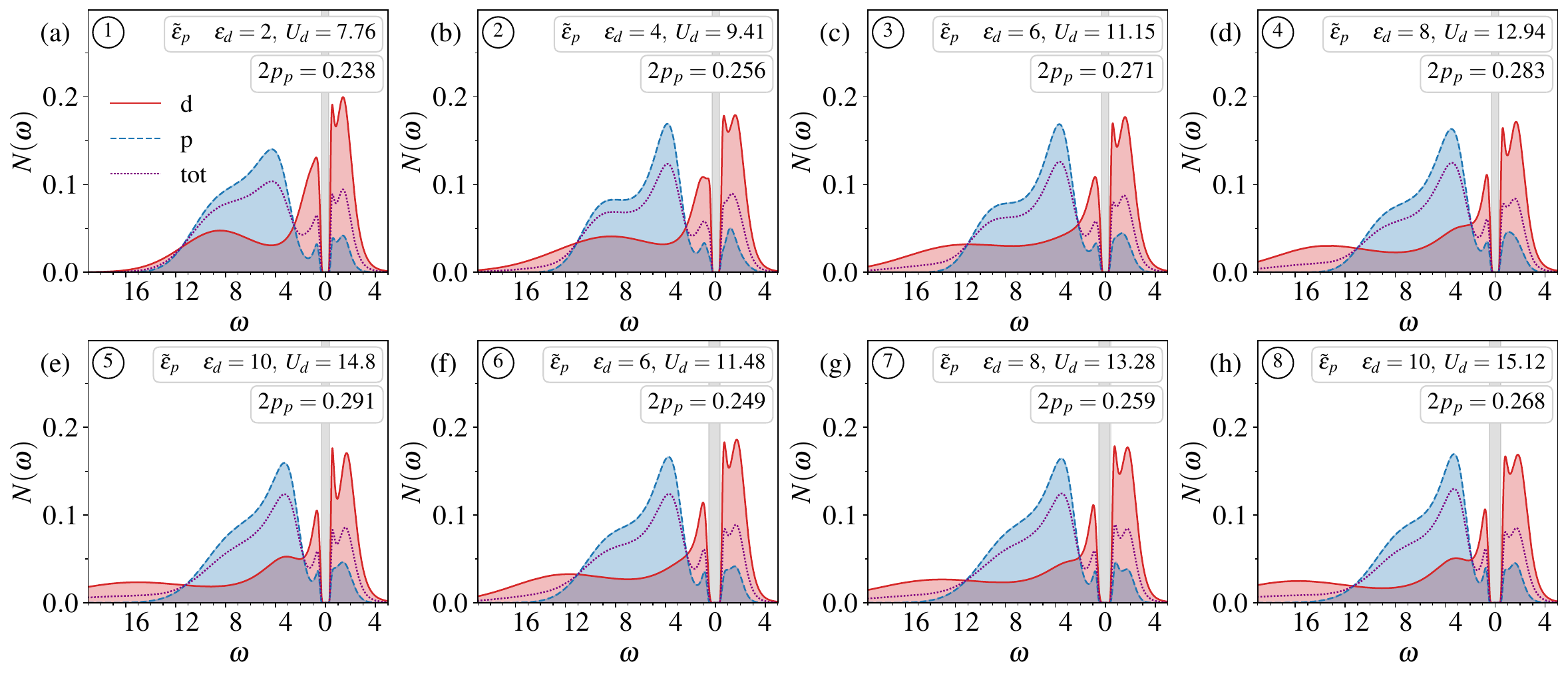}
}
\caption{(a)-(h) Partial density of states $N(\omega)$ of the Cu $3d_{x^2-y^2}$ orbital (red solid line) and of the degenerate O $2p_x$, $2p_y$ orbitals (blue dashed line), along with the total density of states $N_{\rm tot} = [N_d(\omega) +2N_p(\omega)]/3$ (purple dotted line). Each density of states is normalised to unity. The Fermi level is at $\omega=0$. Each panel corresponds to the parameters $(\tilde{\epsilon}_p -\epsilon_d, U_d)$. These parameters are labeled by circled numbers, which are also indicated in the $U_d - \Delta$ diagram of  Fig.~\ref{fig:zsa}(a). Data are for $T=1/50$. We perform the analytical continuation from imaginary to real frequencies with the software of Ref.~\cite{DominicMEM}. Shaded gray region denotes the charge gap size extracted from $N_{\rm tot}(\omega)$ by calculating when $| dN_{\rm tot}(\omega)/d\omega | < 0.1$ around the Fermi level. It agrees with that computed by the width of the plateau at $n_{\rm tot}(\mu)=5$. The value of O hole content $2p_p$ computed from the O partial occupancy is also shown. 
}
\label{fig:dosCTI}
\end{figure*}

First, let us study the charge gap size, denoted by $\Delta_{\rm I}$. We obtain $\Delta_{\rm I}$ from the width of the plateau at $n_{\rm tot}(\mu)=5$. Fig.~\ref{fig:zsa}(b) shows the isotherm $\delta(\mu)= 5 – n_{\rm tot}(\mu)$ for a representative point in the $U_d-\Delta$ diagram. Numerically, we define the plateau at $n_{\rm tot}(\mu)=5$ by the condition $dn_{\rm tot}/d\mu < 0.005$. A plateau in the isotherm $\delta(\mu)$ signals an incompressible insulating state. The plateau occurs when the total number of electrons in the CuO$_2$ unit cell is 5, which is an odd number, thus indicating a correlated, not a band, insulator. The charge gap size $\Delta_{\rm I}$ is a proxy for the spectral gap size between the charge-transfer band and the upper Hubbard band in the local density of states (see Fig.~\ref{fig:dosCTI}), but avoids relying on the analytical continuation.

Tracking the value of $\Delta_{\rm I}$ for each point in the $U_d-\Delta$ phase diagram allows us to determine (i) the charge-transfer insulator to metal boundary (hatched region bounded by thick black lines) and (ii) lines of constant charge gap size (thin black lines). To do that, we compute in Fig.~\ref{fig:zsa}(c) $\Delta_{\rm I}$ as a function of $U_d$ for different values of the bare Cu-O energy distance $\tilde{\epsilon}_p -\epsilon_d$ (i.e. along the dashed gray lines of Fig.~\ref{fig:zsa}(a)). We obtain lines at constant charge gap size in the $U_d-\Delta$ diagram of Fig.~\ref{fig:zsa}(a) as follows: we find the value of $U_d$ that yields a selected $\Delta_{\rm I}$ in the $\Delta_{\rm I}(U_d)$ curve for each value of $\tilde{\epsilon}_p -\epsilon_d$ (i.e., where the horizontal lines in Fig.~\ref{fig:zsa}(c) intercept the $\Delta_{\rm I}(U_d)$ curves). The resulting lines for $\Delta_{\rm I} = 0.6, 0.9, 1.2$ are shown in the $U_d-\Delta$ diagram of Fig.~\ref{fig:zsa}(a). We can also extrapolate $\Delta_{\rm I}(U_d)$ to zero (gray lines in Fig.~\ref{fig:zsa}(c)) to obtain the critical value of $U_d$ for closing the charge gap and thus an estimate for the charge-transfer insulator to metal boundary (dashed black line in Fig.~\ref{fig:zsa}(a) and Fig.~\ref{fig:zsa}(c)). 

However, the true metal-insulator phase boundary is first-order, displaying hysteresis when sweeping $U_d$ up and down. This was demonstrated in Ref.~\cite{Lorenzo3band} for $\tilde{\epsilon}_p -\epsilon_d =7$ with the same method used here (see also Refs.~\cite{go, DeMedici2009}). The inset of Fig.~\ref{fig:zsa}(c) shows the hysteresis loop for $\tilde{\epsilon}_p -\epsilon_d =10$, where the insulating branch (circles) jumps at $U_{d, c1}$ and the metallic branch (squares) jumps at $U_{d, c2}$, and highlights the coexistence region (shaded area). Determining the first-order transition is computationally expensive. Thus, we perform this calculation for $\tilde{\epsilon}_p -\epsilon_d =10$ only. As a result, in the $U_d-\Delta$ diagram of Fig.~\ref{fig:zsa}(a), the spinodal lines $U_{d, c1}(\Delta)$ and $U_{d, c2}(\Delta)$ and the corresponding coexistence region (thick black lines with a hatched area between them) are an estimate of the metal-insulator transition boundary. 

In Fig.~\ref{fig:zsa}(a), the lines of constant $\Delta_{\rm I}$ are nearly parallel and uniformly spaced, indicating a uniform change of charge gap size across the Zaanen-Sawatzky-Allen diagram.

\subsubsection{Degree of charge shared among copper and oxygen orbitals}

Second, let us discuss the degree of charge shared among the copper and oxygen orbitals. The charge-transfer insulating phase occurs at $n_{\rm tot}=n_d+2n_p = 5$, or equivalently at one hole per CuO$_2$ unit cell. Due to the Cu-O hopping amplitude $t_{pd}$, the copper orbital is slightly more than half-filled and the oxygen orbital is almost fully filled, i.e. the partial occupancies are $n_d = 1+2\eta$ and $n_p = 2-\eta$ (see e.g. inset of Fig.~\ref{fig:zsa}(b)). In terms of holes, one can define the copper hole content as $p_d= 2-n_d$ and the oxygen hole content as $2p_p = 2(2-n_p)$, where the extra factor of 2 includes the $2p_x$, $2p_y$ orbital degeneracy. Physically, due to the Cu-O hopping amplitude $t_{pd}$, the hole that localizes in the charge-transfer insulator has a mixed $d$-$p$ character, i.e. is shared between the Cu and O orbitals. The parameter $\eta$ (or any partial occupancy or partial hole content $n_d$, $n_p$, $p_d$, $p_p$) quantifies the mixed $d$-$p$ character of this hole. To connect with experiments in cuprates~\cite{Jurkutat:PRB2014, Rybicki:NatComm2016, Jurkutat:PNAS2023}, we use $2p_p$ as a measure of the degree of charge shared among copper and oxygen orbitals.

Tracking the value of $2p_p $ for each point of the $U_d-\Delta$ phase diagram enables us to determine lines of constant $2p_p$ (pale blue lines). To do that, we compute in Fig.~\ref{fig:zsa}(d) $\Delta_{\rm I}$ as a function of $2p_p$ for different values of the bare Cu-O energy distance $\tilde{\epsilon}_p -\epsilon_d$ (i.e. along the gray lines of Fig.~\ref{fig:zsa}(a)). We obtain lines at constant $2p_p$ in the $U_d-\Delta$ diagram of Fig.~\ref{fig:zsa}(a) as follows: we find the value of $\Delta_{\rm I}$ that yields a selected $2p_p$ (vertical lines in Fig.~\ref{fig:zsa}(d)) in the $\Delta_{\rm I}(2p_p)$ curve for each Cu-O energy distance. The resulting lines for $2p_p \in [0.24, 0.28]$ at a step of $0.1$ are shown in the $U_d-\Delta$ diagram of Fig.~\ref{fig:zsa}(a). They are almost vertical, nearly parallel and uniformly spaced, signalling a dominant change of O hole content upon decreasing $\Delta$. 

\subsubsection{Discussion}

By inspection of Fig.~\ref{fig:zsa}(a), the lines of constant charge gap size $\Delta_{\rm I}$ and lines of constant O hole content $2p_p$ are not parallel. This is significant, as it shows that the physical quantities $\Delta_{\rm I}$ and $2p_p$ are not proportional. Note that if $U_d$ is kept constant, increasing the charge gap size leads to a decrease of $2p_p$, as the lines of constant $2p_p$ are almost vertical. However, in general, there is not a proportionality between $\Delta_{\rm I}$ and $2p_p$. 

This implication was not addressed in Ref.~\cite{GemmaPRB2025} and has important consequences for the optimization of $T_c^{\rm max}$, as we shall discuss in Sec.~\ref{sec:trends}. For now, it is sufficient to note that prior work~\cite{Nicolas:PNAS2021} established that the distance from the metal to insulator boundary determines $T_c^{\rm max}$, i.e. the charge gap size $\Delta_{\rm I}$ controls $T_c^{\rm max}$. However, since the same size of the charge gap can be obtained with different values of $2p_p$ (see Fig.~\ref{fig:zsa}(a)), and since doping redistributes the charge among Cu and O orbitals, this raises the question of the dependence of $T_c^{\rm max}$ on O hole content for a given value of the charge gap size in the parent insulating state. This is the fundamental physical question that we answered in the companion paper~\cite{Eleanor_short}. 

There, we examined the superconducting state emerging from hole doping the insulators at points labeled $1$ to $5$ in Fig.~\ref{fig:zsa}(a), which have equal charge gap size $\Delta_{\rm I} =0.6$, and showed that $T_c^{\rm max}$ increases with increasing O hole content, i.e. in going from $1$ to $5$. Here, we build upon this work by extending the analysis to the superconducting state emerging from hole doping the insulators at points labeled $6$ to $8$ in Fig.~\ref{fig:zsa}(a), which have a larger charge gap size $\Delta_{\rm I} =0.9$. Crucially, this allows us to investigate the interplay between the dependence of $T_c^{\rm max}$ on both O hole content and $\Delta_{\rm I}$. In other words, $\Delta_{\rm I}$ and O hole content are two mechanisms for controlling $T_c^{\rm max}$. Resolving the relationship of $T_c^{\rm max}$ with O hole content  and $\Delta_{\rm I}$ is the focus of Sec.~\ref{sec:trends}. 

Note that the main findings of this section, i.e. the shape of the Zaanen-Sawatzky-Allen diagram of Fig.~\ref{fig:zsa}(a) and the lack of proportionality between $\Delta_{\rm I}$ and $2p_p$ in the parent insulating state, can be tested with recent proposals to realize the Emery model with ultracold atoms~\cite{Lange2026, McCabe2026}.

\subsection{Local density of states}

The Zaanen-Sawatzky-Allen diagram of Fig.~\ref{fig:zsa}(a) reveals that the same charge gap size can be realized with different values of the bare charge-transfer energy $\Delta$ (or the bare Cu-O energy distance $\tilde{\epsilon}_p -\epsilon_d$), and gives different mixed $d$-$p$ character of the bands, as indicated by different values of the O hole content. 

To further characterize these results, Fig.~\ref{fig:dosCTI} shows the partial density of states on copper (red) and oxygen (blue) orbitals, along with the total density of states (purple). Data correspond to the charge-transfer insulators at points labeled $1$ to $8$ in the Zaanen-Sawatzky-Allen diagram of Fig.~\ref{fig:zsa}(a). Insulators at points $1$ to $5$ have charge gap size $\Delta_{\rm I} =0.6$, whereas those at points $6$ to $8$ have charge gap size $\Delta_{\rm I} =0.9$. 
The insulating gap occurs between the charge-transfer band and the upper Hubbard band. The degree of the mixed $d$-$p$ character, quantified by $2p_p$, increases in going from the insulator at point $1$ to the insulator at point $5$ (or from $6$ to $8$). Physically, this is expected: the mixed $d$-$p$ character increases as we move deeper into the charge-transfer regime (region where $U_d \gg \Delta$ in Fig.~\ref{fig:zsa}(a), or far away from the Mott-Hubbard regime (region where $U_d < \Delta$ in Fig.~\ref{fig:zsa}(a)).

\section{Superconductivity in hole-doped charge-transfer insulators}
\label{sec:SC}

This section studies the superconducting state emerging from doping with holes the charge-transfer insulators of Fig.~\ref{fig:dosCTI}, i.e. the insulators at points with labels $1$ to $8$ in Fig.~\ref{fig:zsa}(a). We analyze the superconducting critical temperature dependence on hole doping $\delta$ and on O hole content $2p_p$, and then we characterize the superconducting state with the local density of states.

\subsection{Superconducting dome vs hole doping}

\begin{figure*}[t!]
\centering{
\includegraphics[width=1.0\linewidth]{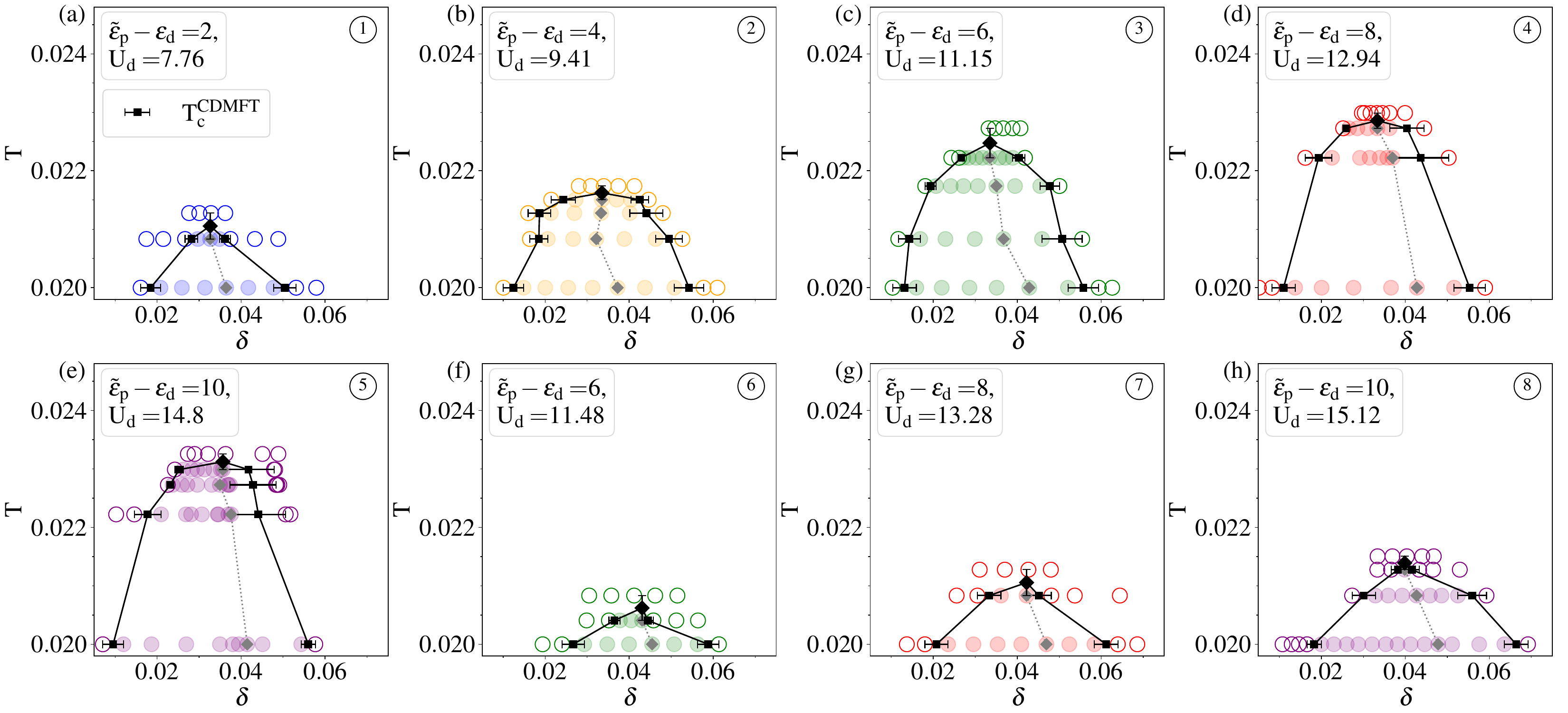}
}
\caption{(a)-(h) Superconducting critical temperature $T_c^{\rm CDMFT}$ (black line with filled squares) vs hole doping $\delta=5-n_{\rm tot}$. $T_c^{\rm CDMFT}(\delta)$ is the line where the superconducting order parameter $\Phi$ changes from zero (open circles) to nonzero (shaded circles) [as determined in Fig.~\ref{fig:SCopvsdoping}]. The dotted gray line with diamonds connects the values of doping that maximize $\Phi$ at each temperature. A black diamond marks the maximum superconducting critical temperature $T_c^{\rm max}$. Each panel corresponds to the parameters $(\tilde{\epsilon}_p -\epsilon_d, U_d)$. These parameters are labeled by circled numbers. The parent ($\delta=0$) insulating state from which the superconducting state emerges is indicated in the Zaanen-Sawatzky-Allen diagram of Fig.~\ref{fig:zsa}(a).
}
\label{fig:Tcvsdoping}
\end{figure*}
\begin{figure*}[t!]
\centering{\includegraphics[width=1.0\linewidth]{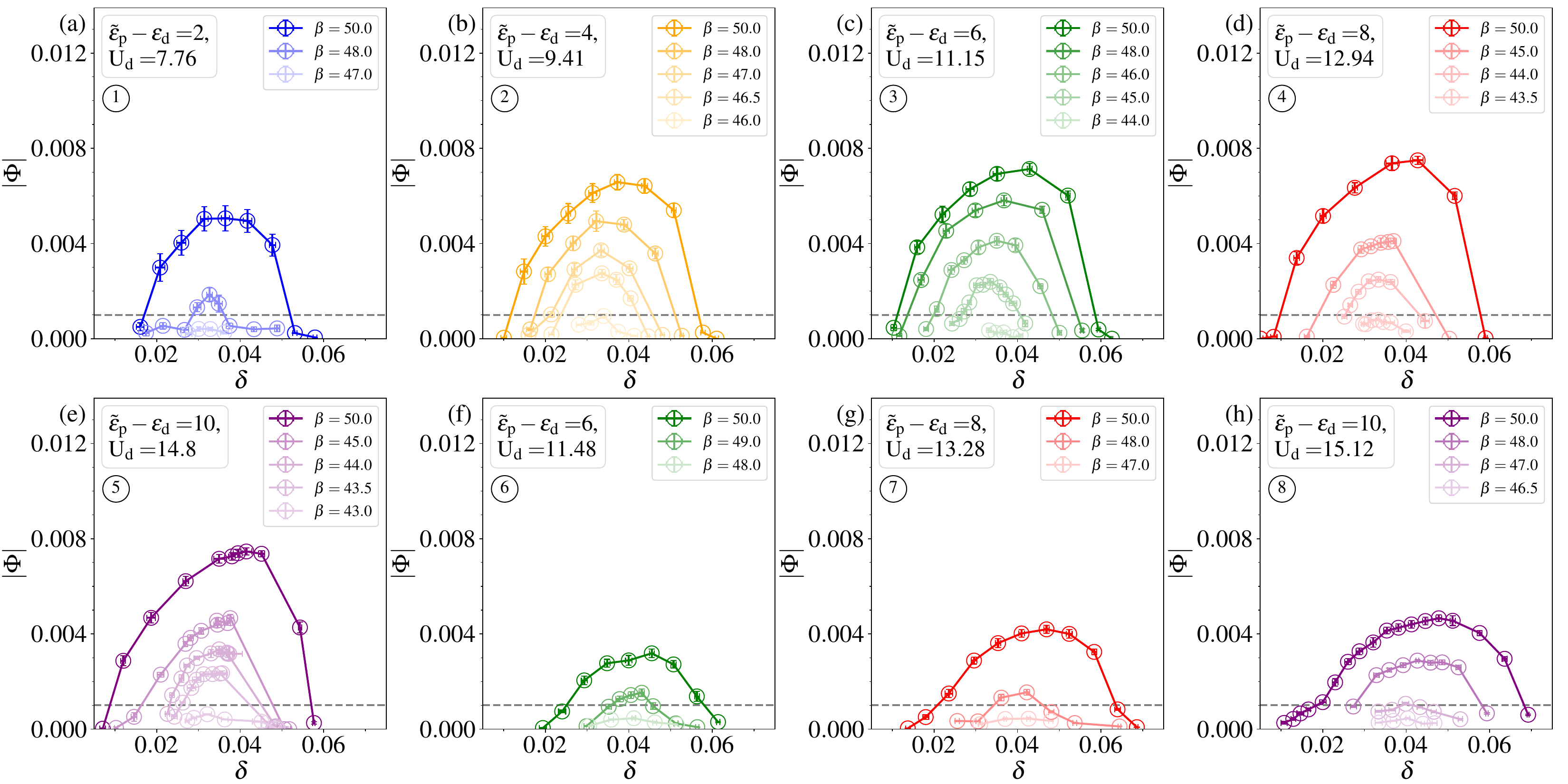}}
\caption{(a)-(h) Superconducting order parameter $\Phi$ vs doping $\delta$ for different values of the inverse temperature $\beta=1/T$. Each panel corresponds to coordinates $(\tilde{\epsilon}_p -\epsilon_d, U_d)$. The horizontal dashed line at $\Phi=0.001$ marks the threshold above which we consider the system to be superconducting.
}
\label{fig:SCopvsdoping}
\end{figure*}

Figure~\ref{fig:Tcvsdoping} shows the superconducting critical temperature $T_c^{\rm CDMFT}$ versus hole doping $\delta$ emerging from the parent charge-transfer insulating states at points labeled $1$ to $8$ in the $U_d - \Delta$ diagram of Fig.~\ref{fig:zsa}(a). Hole doping is defined as $\delta=5-n_{\rm tot}$. $T_c^{\rm CDMFT}$ is the temperature below which the superconducting order parameter $\Phi$ is nonzero (numerically, we take $|\Phi| > 0.001$), where $\Phi = \pm \langle d_i^\dagger d_j^\dagger \rangle$ with $i$ and $j$ as nearest neighbor Cu sites, and where the sign changes upon $\pi/2$ rotation. Shaded circles mark points where the system is superconducting and open circles where it is not. Figure~\ref{fig:SCopvsdoping} shows the superconducting order parameter $\Phi$ versus $\delta$ for the same model parameters of Fig.~\ref{fig:Tcvsdoping}. For each temperature, the value of doping that maximizes $\Phi(\delta)$ is shown with a gray diamond in Fig.~\ref{fig:Tcvsdoping}.

Despite thermal fluctuations destroying long-range order in two dimensions~\cite{MWtheorem}, $T_c^{\rm CDMFT}$ is a mean-field temperature and physically denotes when superconducting pairs develop within the cluster. 

In Fig.~\ref{fig:Tcvsdoping}(a)-(h), the superconducting state shows a dome-like shape versus $\delta$~\cite{Kent:PRB2008, ArrigoniCuO2, Weber2011, Lorenzo3band, Dash:PRB2019, Cui:PRR2020, Nicolas:PNAS2021, Mai:PRB2021, Mai:npj2021, Ponsioen:PRB2023, St-Cyr:2025, Jacob:2026}, reaching its maximum value $T_c^{\rm max}$ at a finite doping (black diamond) which is dependent on the bare charge-transfer energy $\Delta$. The behavior of $T_c^{\rm max}$ with $\Delta$ will be analyzed in Sec.~\ref{sec:trends}. 
The Monte-Carlo sign problem prevents us from reaching the low temperatures (see Appendix~\ref{sec:sign}). Hence we can only calculate the upper part of the superconducting dome (note that the $y$ axis in Fig.~\ref{fig:Tcvsdoping} does not start at zero). 
The superconducting state occurs in a wider region of doping with decreasing temperature, as indicated by CDMFT calculations at zero-temperature~\cite{Dash:PRB2019, Nicolas:PNAS2021, St-Cyr:2025}. For example, for $\tilde{\epsilon}_p -\epsilon_d=7$ and $U_d=12$, Ref.~\cite{Dash:PRB2019} shows that the zero-temperature width of the superconducting dome extends up to $\delta\approx 0.14$.
  
$T_c^{\rm CDMFT}(\delta)$ is not proportional to superconducting order parameter at the lowest temperature (here, $T=1/50$) $\Phi(\delta)|_{T=1/50}$. In particular, optimal doping, i.e. the doping that maximizes $T_c^{\rm CDMFT}(\delta)$, does not in general correspond to the doping that optimizes $\Phi(\delta)|_{T=1/50}$ (in other words, in Fig.~\ref{fig:Tcvsdoping} the doping of the black diamond does not correspond to the doping of the gray diamond at the lowest temperature). As a result, the superconducting dome is slightly asymmetric as a function of doping and the locus of $\Phi$ maxima in the $T-\delta$ plane traces a negatively sloped line (dotted line with gray diamonds). This non-BCS behavior also occurs in the two-dimensional single-band Hubbard model~\cite{LorenzoSC}.

\subsection{Superconducting dome vs oxygen hole content}

\begin{figure}[t!]
\centering{
\includegraphics[width=1.0\linewidth]{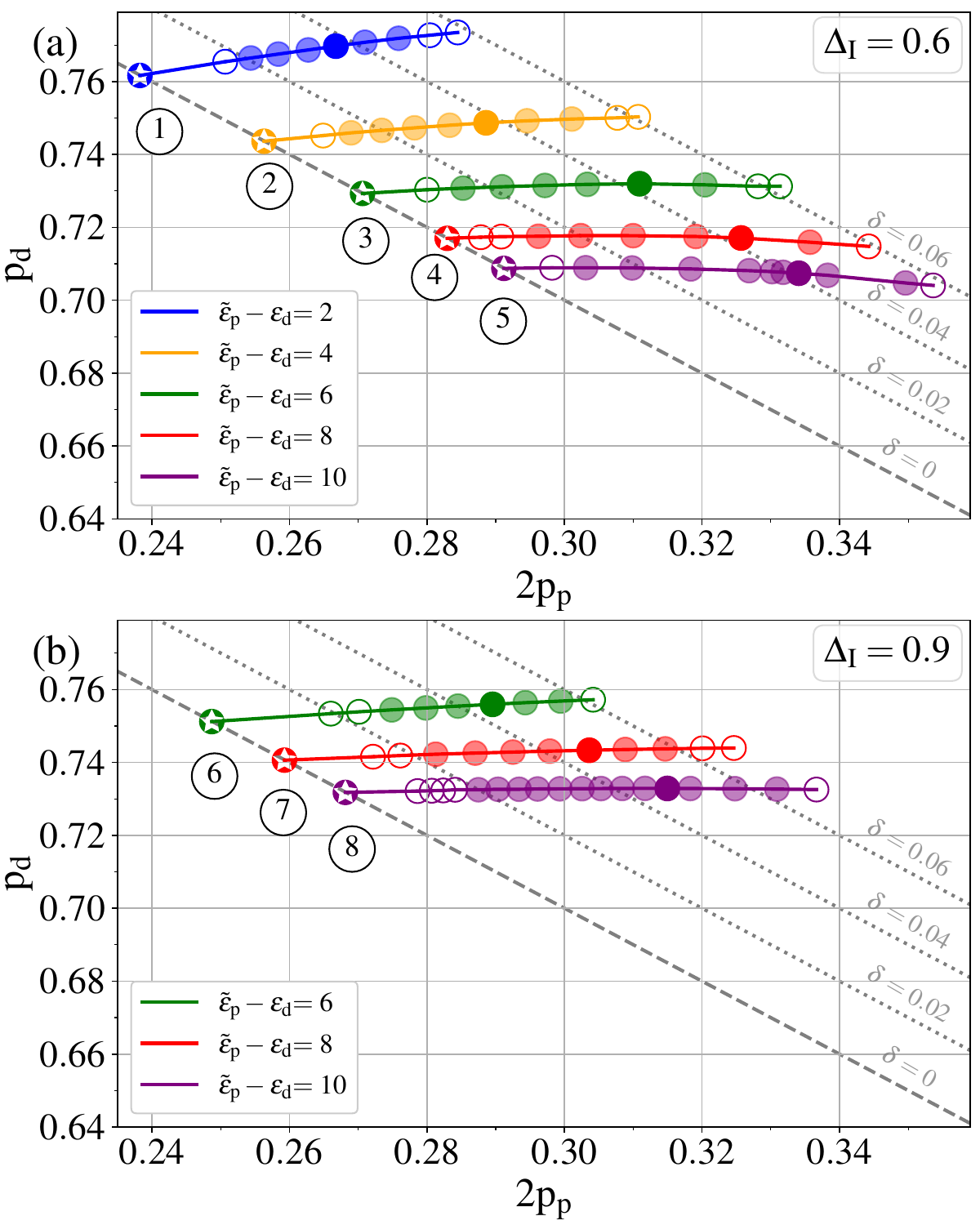}
}
\caption{Copper hole content $p_d$ vs oxygen hole content $2p_p$ for $T=1/50$ and charge gap size $\Delta_{\rm I}=0.6$ (panel(a)) and $\Delta_{\rm I}=0.9$ (panel (b)). Same color code as in Fig.~\ref{fig:zsa}. In each panel, the gray dashed line indicates zero doping, whereas the dotted gray lines indicate constant values of hole doping, at a step of $0.02$. Shaded circles denote the superconducting state and open circles the normal state. Filled solid circles represent where the superconducting order parameter is maximum. Circles with a star mark the parent charge-transfer insulating state at zero doping. These states are labeled by circled numbers and are also indicated in Fig.~\ref{fig:zsa}(a).
}
\label{fig:CHOvsOHC}
\end{figure}

We established that superconductivity emerges upon doping charge-transfer insulators. Upon doping, the electronic charge is also redistributed among Cu and O orbitals~\cite{Jurkutat:PRB2014, Rybicki:NatComm2016, Jurkutat:PNAS2023}. To quantify this charge redistribution, Fig.~\ref{fig:CHOvsOHC} shows the Cu hole content $p_d$ vs O hole content $2p_p$ in the superconducting state emerging from the parent charge-transfer insulating states labeled $1$ to $8$ in Fig.~\ref{fig:zsa}(a), for the low temperature $T=1/50$. As in Fig.~\ref{fig:Tcvsdoping}, shaded circles indicate the superconducting state and open circles the normal state, with the same color code of Fig.~\ref{fig:zsa}(a). 

The diagonal dashed gray line marks zero doping. Data on this line represent charge-transfer insulators and are indicated by circles with a star, which correspond to the same symbols with labels $1$ to $8$ in Fig.~\ref{fig:zsa}(a).
The dotted gray lines parallel to the $\delta=0$ line indicate constant values of hole doping. Doping increases as we move away from the $\delta=0$ line. The effect of hole doping is to redistribute the hole content among Cu and O orbitals. The small slope of the curves indicate that hole doping mostly increases the O hole content with a small effect on the Cu hole content~\cite{Jurkutat:PRB2014, Rybicki:NatComm2016, Jurkutat:PNAS2023, Scalettar:PRB1991, White:PRB2015, Lorenzo3band, Kung:PRB2016, Dash:PRB2019, Nicolas:PNAS2021, GiovanniPRB2025, St-Cyr:2025, GemmaPRB2025, Peng:PRB2025, Polat:2026}. Physically, this reveals that the doped holes mostly enter the O orbitals~\cite{zsa, Emery_1987}, an effect beyond the physics of the single-band Hubbard model.  
Upon decreasing the bare charge-transfer energy $\Delta$ (i.e. increasing simultaneously $\tilde{\epsilon}_p -\epsilon_d$ and $U_d$ such that we move from points labeled $1$ to $5$ or $6$ to $8$), the curves overall shift toward the bottom right. Physically, this reveals that as the system goes deeper into the charge-transfer regime (i.e. along a given thin black line from bottom right to top left in Fig.~\ref{fig:zsa}(a)), the doped holes enter the O orbitals even more favorably.

\begin{figure*}[t!]
\centering{
\includegraphics[width=1.0\linewidth]{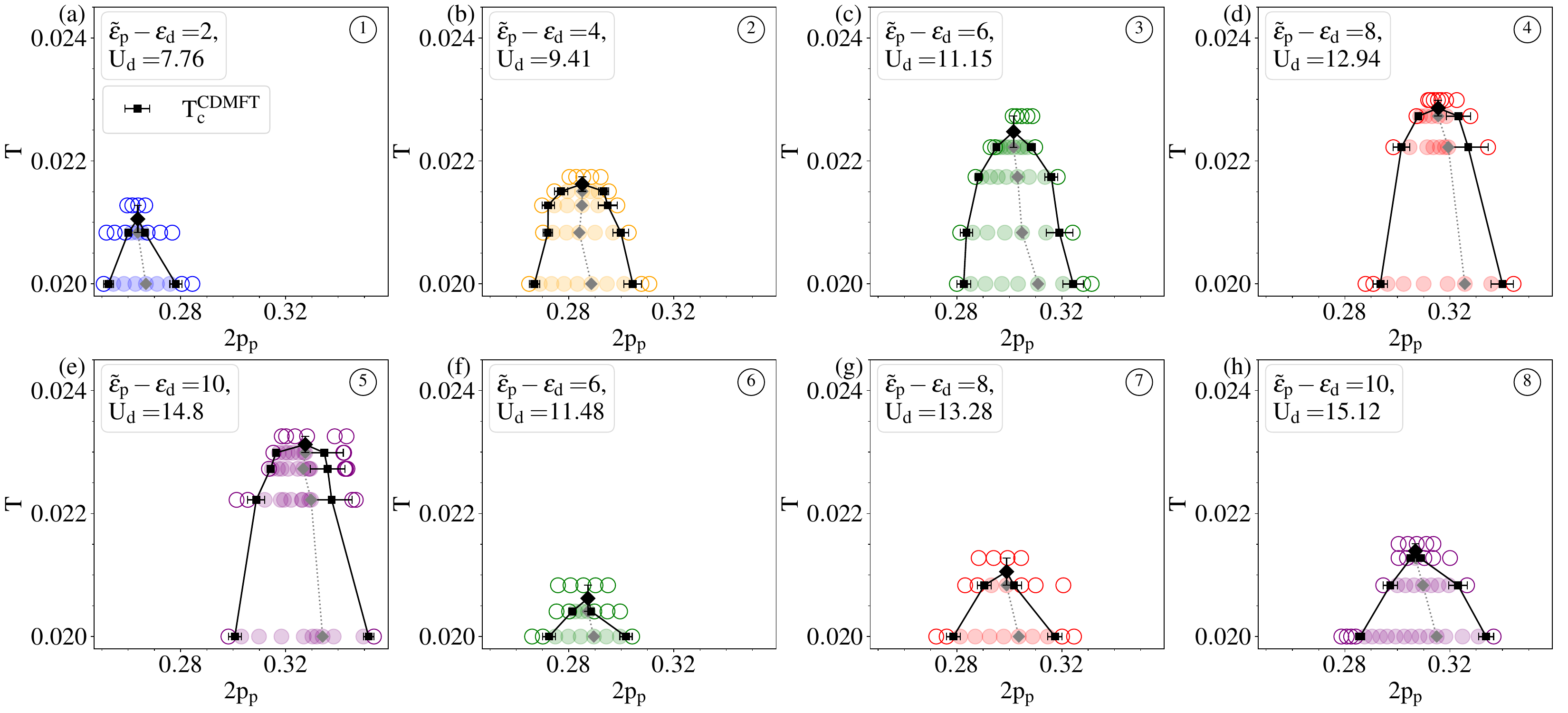}
}
\caption{(a)-(h) Same as Fig.~\ref{fig:Tcvsdoping}, but with oxygen hole content $2p_p$ instead of hole doping $\delta$ on the $x$-axis.}
\label{fig:TcvsOHC}
\end{figure*}

Having obtained the O hole content, we can now follow the experimental Refs.~\cite{Rybicki:NatComm2016, Jurkutat:PNAS2023} and plot the superconducting critical temperature as a function of O hole content $2p_p$ rather than hole doping $\delta$. Fig.~\ref{fig:TcvsOHC}(a)-(h) shows $T_c^{\rm CDMFT}$ versus O hole content that emerges from the parent charge-transfer insulating states labeled $1$ to $8$ in Fig.~\ref{fig:zsa}(a). 
Note that Fig.~\ref{fig:Tcvsdoping}(a)-(h) and Fig.~\ref{fig:TcvsOHC}(a)-(h) show the same data using different $x$ axes. The superconducting state shows a dome-like shape not only as a function of doping $\delta$ but as a function of O hole content too. A black diamond indicates the coordinates of optimal O hole content and $T_c^{\rm max}$. The behavior of $T_c^{\rm max}$ with $2p_p$ will be examined in Sec.~\ref{sec:trends}.

\subsection{Local density of states}

\begin{figure*}[t!]
\centering{
\includegraphics[width=1.0\linewidth]{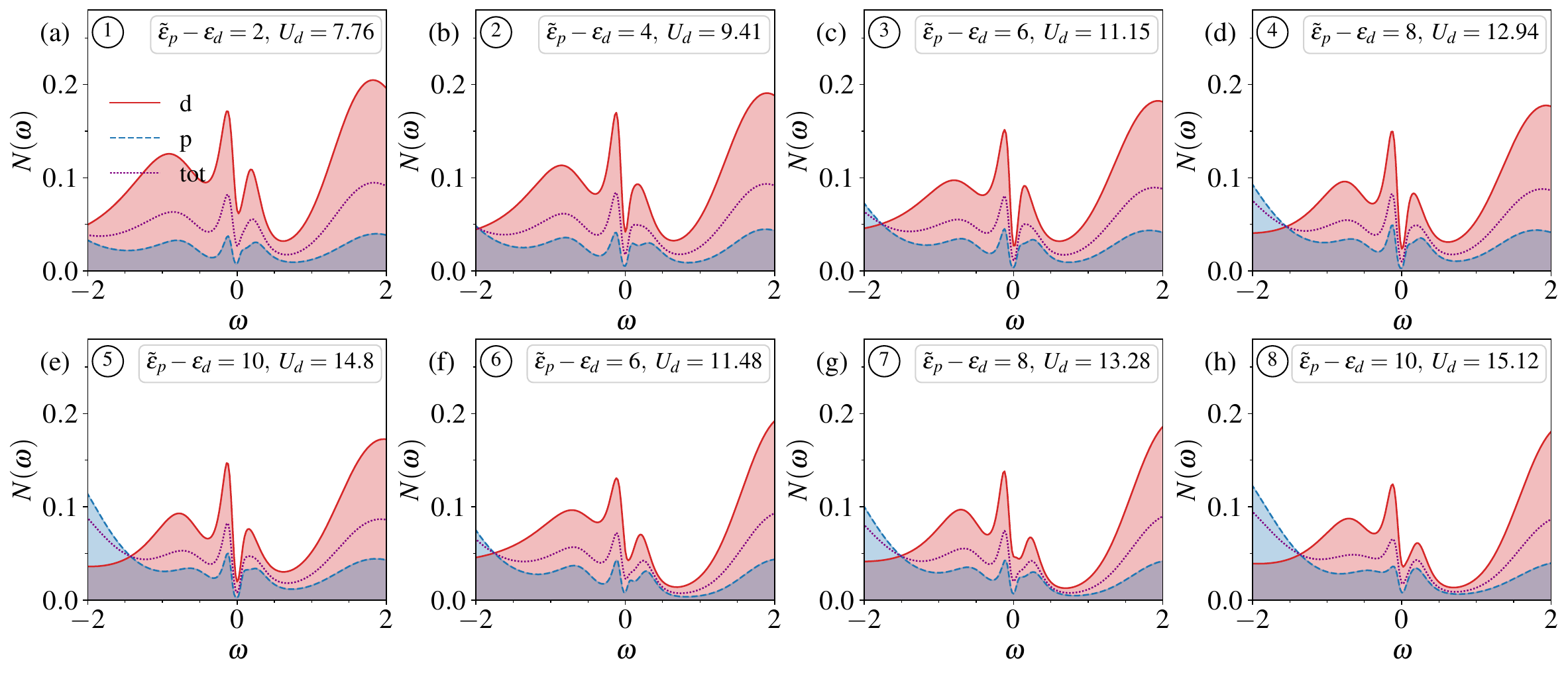}
}
\caption{(a)-(h) Same as Fig.~\ref{fig:dosCTI}, but in the superconducting state obtained upon hole doping the parent charge-transfer insulators of Fig.~\ref{fig:dosCTI}. Data are at the small doping $\delta \approx 0.03$ and at $T=1/50$.
}
\label{fig:dosSC}
\end{figure*}

Next we characterize the superconducting state emerging from the parent charge-transfer insulating states labeled $1$ to $8$ in Fig.~\ref{fig:zsa}(a) with the local density of states. Figure~\ref{fig:dosSC} shows the partial density of states on copper (red) and oxygen (blue) orbitals, as well as the total density of states (purple). Data are taken at our lowest temperature, $T=1/50$, and at the low doping $\delta \approx 0.03$. The doping dependence of the local density of states is analyzed in Refs.~\cite{Lorenzo3band, Dash:PRB2019} for $\tilde{\epsilon}_p -\epsilon_d=7$. 

The spectra show superconducting coherence peaks both in the copper and oxygen partial density of states, indicating that superconducting pairs are composite objects of mixed $d$-$p$ character. The degree of the mixed $d$-$p$ character, quantified by $2p_p$, increases in going from $1$ to $5$ or from $6$ to $8$ (see Fig.~\ref{fig:CHOvsOHC}).

\section{Trends in $T_c^{\rm max}$}
\label{sec:trends}

The preceding Sec.~\ref{sec:SC} showed that a superconducting dome emerges upon hole doping a charge-transfer insulator at different points of the Zaanen-Sawatzky-Allen diagram. The focus of this section is to examine the trends of the maximum value of the superconducting dome, $T_c^{\rm max}$, as a function of model parameters $\Delta$ and $U_d$, and then as a function of physical quantities $\Delta_{\rm I}$ and $2p_p$.

\subsection{$T_c^{\rm max}$ vs model parameters $\Delta$ and $U_d$}

First, we examine $T_c^{\rm max}$ in the Zaanen-Sawatzky-Allen scheme, i.e. as a function of the model parameters $\Delta$ and $U_d$.

\subsubsection{$T_c^{\rm max}$ vs bare charge transfer energy $\Delta$ for given values of charge gap size $\Delta_{\rm I}$}

\begin{figure}[t!]
\centering{
\includegraphics[width=1.0\linewidth]{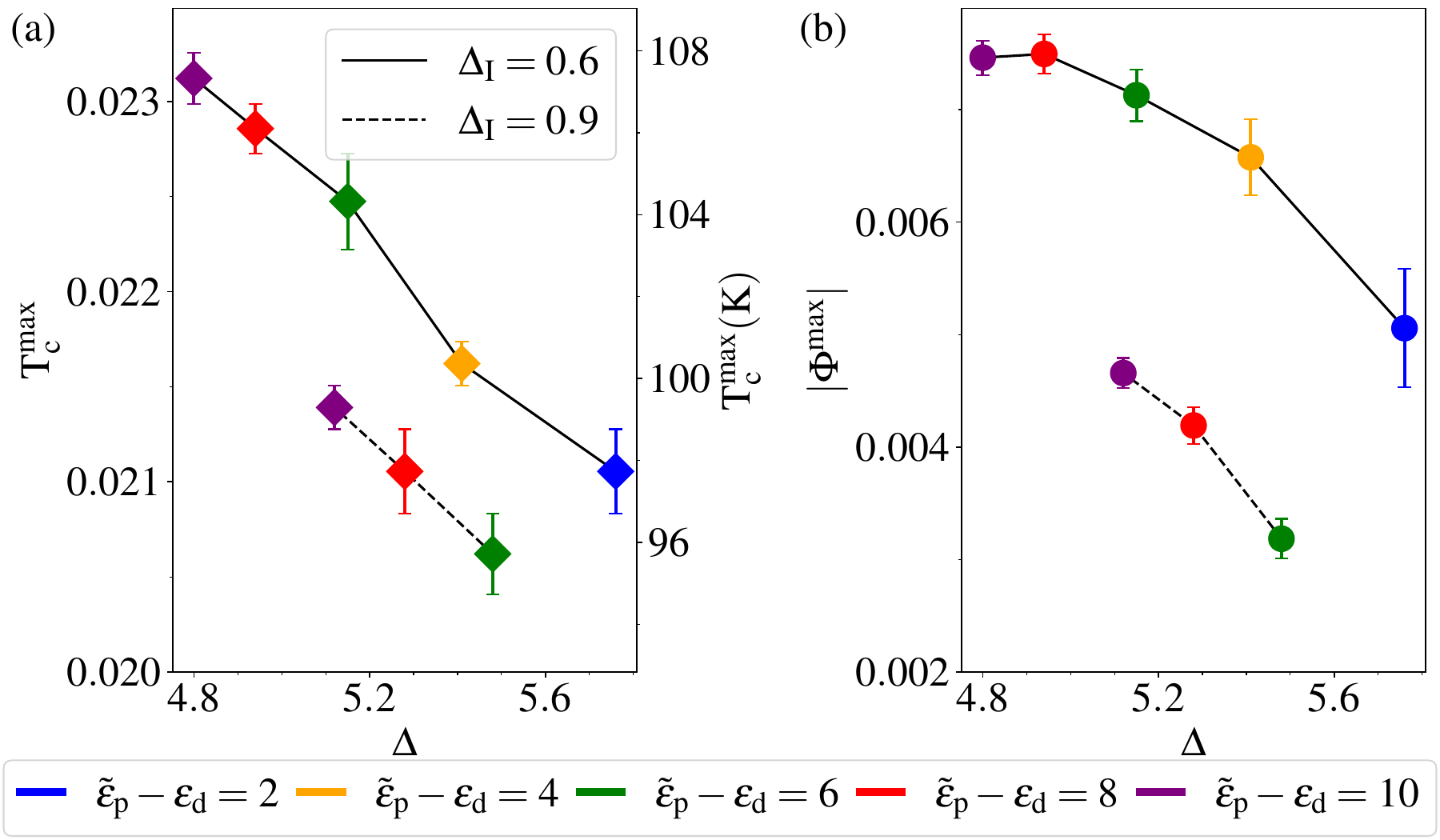}
}
\caption{(a) Maximum superconducting critical temperature $T_c^{\rm max}$ vs bare charge transfer energy $\Delta= \epsilon_d + U_d -\tilde{\epsilon}_p$ for two values of the charge gap size $\Delta_{\rm I}$. Temperature is converted into Kelvin with $t_{pp}=0.4$~eV. (b) Same as (a), but with maximum superconducting order parameter $\Phi^{\rm max}$ at $T=1/50$ instead of $T_c^{\rm max}$ on the $y$ axis. Each data point shows the coordinates $(\Delta, T_c^{\rm max})$ [$(\Delta, \Phi^{\rm max})$] found in each panel of Fig.~\ref{fig:Tcvsdoping} [Fig.~\ref{fig:SCopvsdoping}]. 
}
\label{fig:TcmaxvsDelta}
\end{figure}
\begin{figure}[t!]
\centering{
\includegraphics[width=1.0\linewidth]{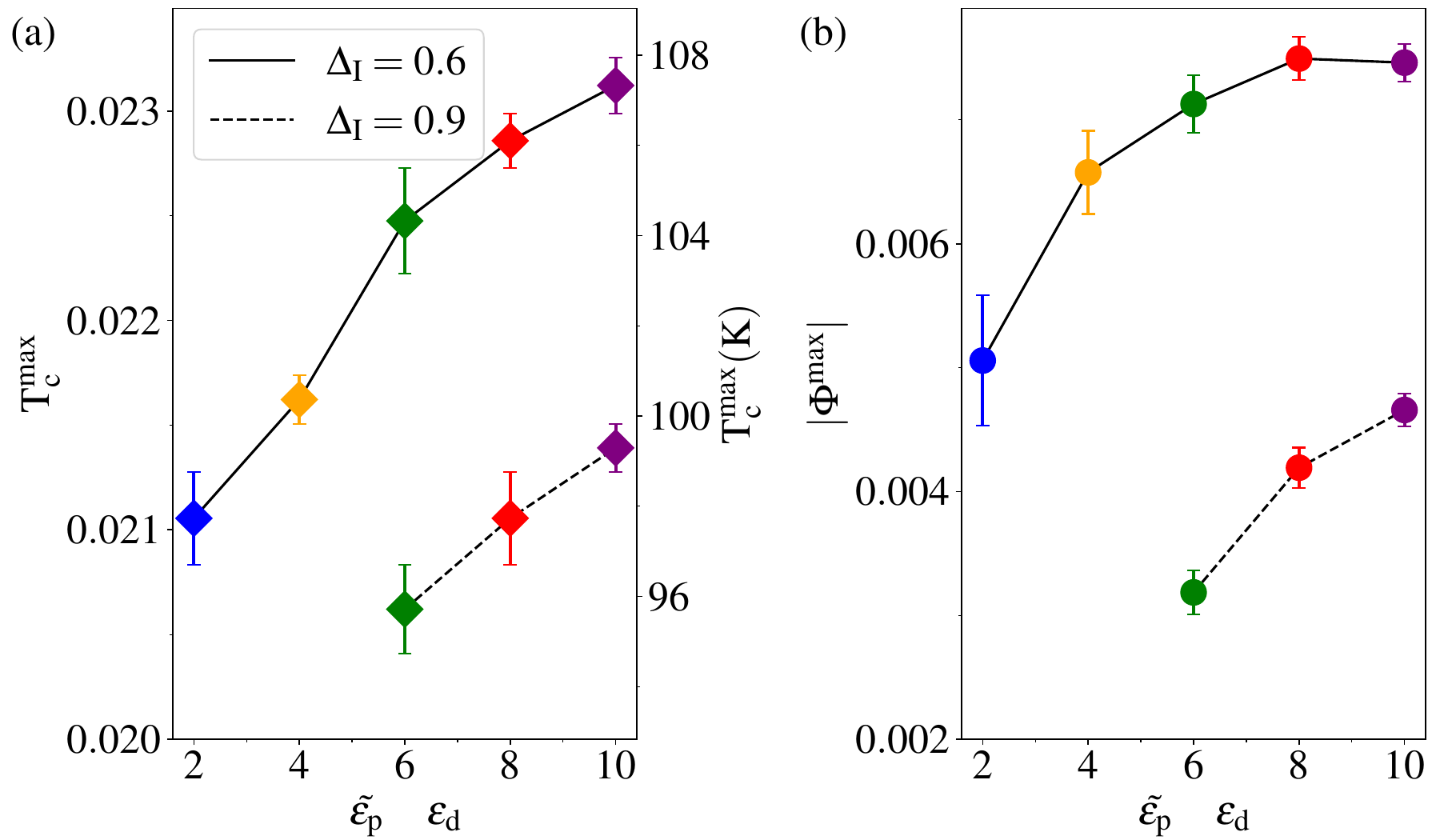}
}
\caption{Same as Fig.~\ref{fig:TcmaxvsDelta}, but vs bare Cu-O energy distance $\tilde{\epsilon}_p -\epsilon_d$. 
}
\label{fig:Tcmaxvsep}
\end{figure}

Figure~\ref{fig:TcmaxvsDelta}(a) [Fig.~\ref{fig:Tcmaxvsep}(a)] shows $T_c^{\rm max}$ as a function of the bare charge-transfer energy $\Delta$ [bare Cu-O energy distance $\tilde{\epsilon}_p -\epsilon_d$] for two given values of the charge gap size $\Delta_{\rm I}$ in the parent ($\delta=0$) state, $\Delta_{\rm I}=0.6$ (solid line) and $\Delta_{\rm I}=0.9$ (dashed line). Each diamond symbol corresponds to the value of $T_c^{\rm max}$ obtained in Fig.~\ref{fig:Tcvsdoping}, with the same color code. In other words, each diamond symbol corresponds to the value of $T_c^{\rm max}$ that occurs upon doping with holes the charge-transfer insulators at points labeled $1$ to $5$ or $6$ to $8$ in Fig.~\ref{fig:zsa}(a). 

For a given $\Delta_{\rm I}$, $T_ c^{\rm max}$ decreases with increasing $\Delta$ (Fig.~\ref{fig:TcmaxvsDelta}(a)). Equivalently,  $T_ c^{\rm max}$ increases with increasing the bare Cu-O energy distance $\tilde{\epsilon}_p -\epsilon_d$ (Fig.~\ref{fig:Tcmaxvsep}(a)). This relationship is one key finding of the companion paper~\cite{Eleanor_short}, where it is reported for $\Delta_{\rm I}=0.6$ only. Here we confirm that this trend is not restricted to small values of $\Delta_{\rm I}$, but persists for larger values of $\Delta_{\rm I}$, as shown by the curve at  $\Delta_{\rm I} =0.9$. As noted in the companion paper~\cite{Eleanor_short}, this result is consistent with the data of Refs.~\cite{Weber2011, Nicolas:PNAS2021} and challenges prior thinking that $T_ c^{\rm max}$ anticorrelates with the bare Cu-O energy distance $\tilde{\epsilon}_p -\epsilon_d$~\cite{Nicolas:PNAS2021}, and opens up new paths for increasing $T_ c^{\rm max}$. Note that here the hopping amplitudes are fixed. 

Although previous studies commonly rely on the superconducting order parameter $\Phi$ as a proxy for $T_c$~\cite{Weber2011, Nicolas:PNAS2021, St-Cyr:2025}, a strength of our work is to directly determine $T_c$, at least at a CDMFT level ($T_c^{\rm CDMFT}$), i.e. neglecting Kosterlitz-Thouless physics. In principle, this allows us to verify if the maximum superconducting order parameter $\Phi^{\rm max}$ is a good proxy for $T_c^{\rm max}$. Fig.~\ref{fig:TcmaxvsDelta}(b) [Fig.~\ref{fig:Tcmaxvsep}(b)] shows $\Phi^{\rm max}$ at our lowest temperature, $T=1/50$, as a function of the bare charge-transfer energy $\Delta$ [bare Cu-O energy distance $\tilde{\epsilon}_p -\epsilon_d$] for two given values of $\Delta_{\rm I}$. Each circle symbol has been extracted from Fig.~\ref{fig:SCopvsdoping}.  $\Phi^{\rm max}$ and $T_c^{\rm max}$ both decrease with increasing $\Delta$, however their rate of decrease differ. We have already noted in Sec.~\ref{sec:SC} that $\Phi(\delta)|_{T=1/50}$ is not in general proportional to $T_c^{\rm CDMFT}(\delta)$. Further data are needed to determine whether $\Phi^{\rm max}$ is a good proxy for  $T_c^{\rm max}$.

Fig.~\ref{fig:Tcmaxvsep}(a) also shows that, for a given bare Cu-O energy distance $\tilde{\epsilon}_p -\epsilon_d$, $T_c^{\rm max}$ decreases with increasing $\Delta_{\rm I}$, and hence with increasing $U_d$ (see diamonds with the same color code). This finding is in agreement with the results of Ref.~\cite{Nicolas:PNAS2021} at $\tilde{\epsilon}_p -\epsilon_d \in [6,7]$ (see also Appendix~\ref{sec:comparison}). The decrease of $T_c^{\rm max}$ with increasing $U_d$ is expected from strongly correlated superconductivity~\cite{AMJulich} and can also be described by the single-band Hubbard model~\cite{Scalapino:1994, AMJulich, sshtSC, Gull:2013, LorenzoSC, QinAnnuRev2022, Eleanor_PRL2026}. Physically, this trend indicates a link between superconductivity and the superexchange mechanism. Over the years, detailed investigations have indeed shown that the superconducting pairing arises from low energy scales ascribed to short-range spin singlet correlations generated by the superexchange interaction~\cite{Scalapino:1994, AMJulich, sshtSC, Gull:2013, LorenzoSC, QinAnnuRev2022, Eleanor_PRL2026}. Since the present work shows that not only the charge-transfer gap but also O hole content controls $T_c^{\rm max}$, an important future direction opened up by our work will be to address the interplay of these two mechanisms for the dynamics of the superconducting pairing, for example by applying the methodology developed in Ref.~\cite{Eleanor_PRL2026} for the Hubbard model to the Emery model.

\subsubsection{Discussion}

\begin{figure}%[t!]
\centering{
\includegraphics[width=1.0\linewidth]{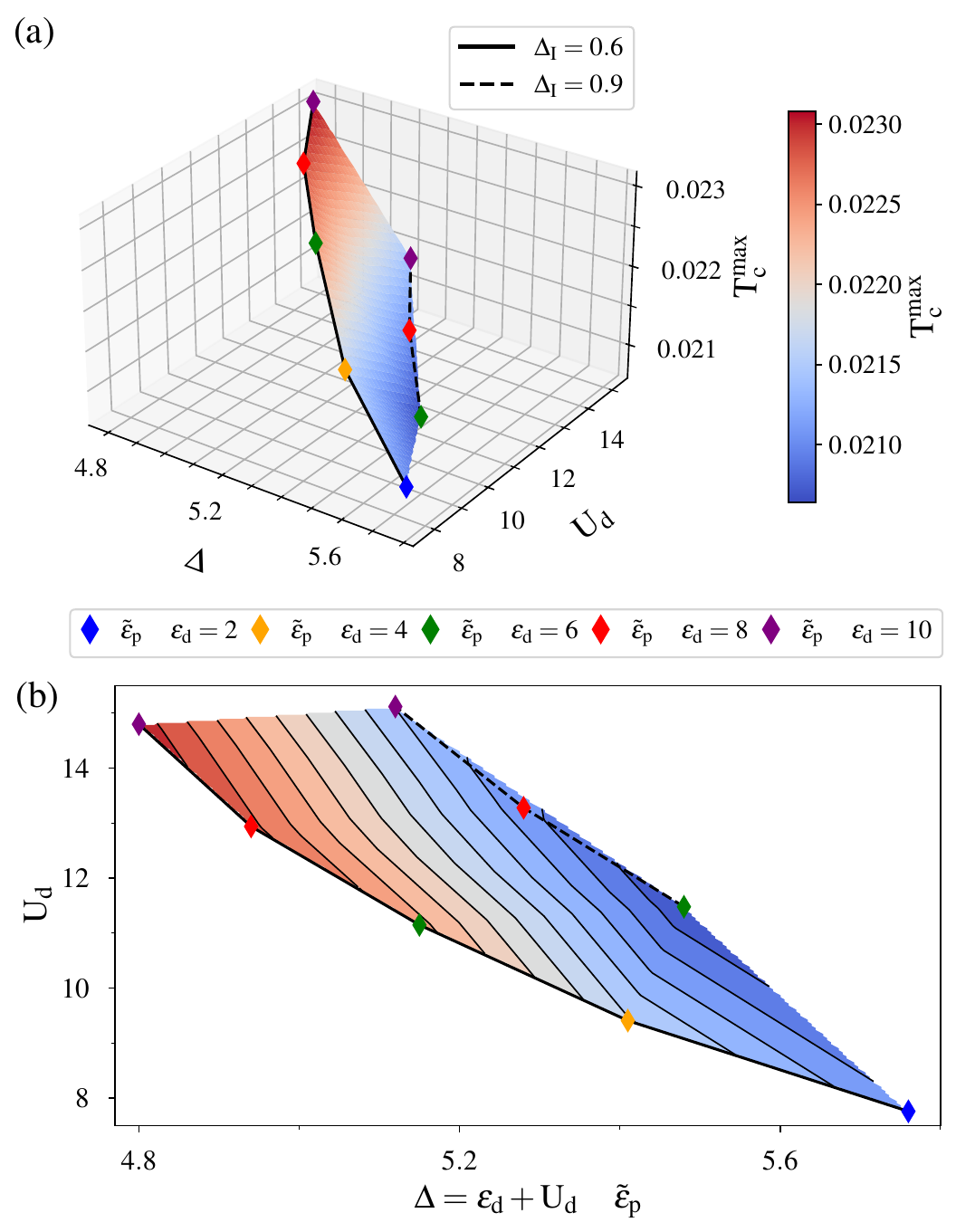}
}
\caption{(a) Maximum superconducting critical temperature $T_c^{\rm max}$ vs bare charge-transfer energy $\Delta$ and interaction strength $U_d$. The surface $T_c^{\rm max}(\Delta, U_d)$ has been obtained by interpolating our data (diamonds) on a uniform grid. The color gradient indicates the magnitude of $T_c^{\rm max}$ and is given in the colorbar. (b) Contour plot of $T_c^{\rm max}$ vs $\Delta$ and $U_d$ with superimposed lines of constant $T_c^{\rm max}$ (black lines). In other words, this panel translates the $T_c^{\rm max}$ trends within the Zaanen-Sawatzky-Allen diagram $U_d$ vs $\Delta$. Note that $x$ and $y$ axes do not have equal aspect ratios. 
}
\label{fig:TcmaxVSDeltaU}
\end{figure}

We identified two key trends: (i) for a fixed charge gap size $\Delta_{\rm I}$, $T_c^{\rm max}$ increases with decreasing the bare charge-transfer $\Delta$ (or with increasing $\tilde{\epsilon}_p -\epsilon_d$); (ii) for a fixed $\tilde{\epsilon}_p -\epsilon_d$, $T_c^{\rm max}$ increases with decreasing $U_d$. 
Let us translate these trends within the Zaanen-Sawatzky-Allen framework. This is done in Fig.~\ref{fig:TcmaxVSDeltaU}, where we show $T_c^{\rm max}$ vs $\Delta$ and $U_d$ (panel (a)), with its contourmap (panel (b)). 
According to the first result, maximizing $T_c^{\rm max}$ requires doping a charge-transfer insulator placed deep in the charge-transfer regime (i.e. region of large $U_d$ and small $\Delta$). According to the second result, maximizing $T_c^{\rm max}$ requires doping a charge-transfer insulator located close to the metal-insulator boundary. The first result is the key finding of the companion article~\cite{Eleanor_short}, whereas the second result was already established in prior work~\cite{Scalapino:1994, Dash:PRB2019, Nicolas:PNAS2021, St-Cyr:2025}. Here, we combine the two findings and map them into the Zaanen-Sawatzky-Allen diagram, as condensed in Fig.~\ref{fig:TcmaxVSDeltaU}. 

Note that the Zaanen-Sawatzky-Allen diagram describes the Emery model in the normal state at zero doping. Hence, it is remarkable that the parent normal state provides an indication of the superconducting critical temperature~\cite{Nicolas:PNAS2021, Wang:Science2023}.

\begin{figure}%[t!]
\centering{
\includegraphics[width=1.0\linewidth]{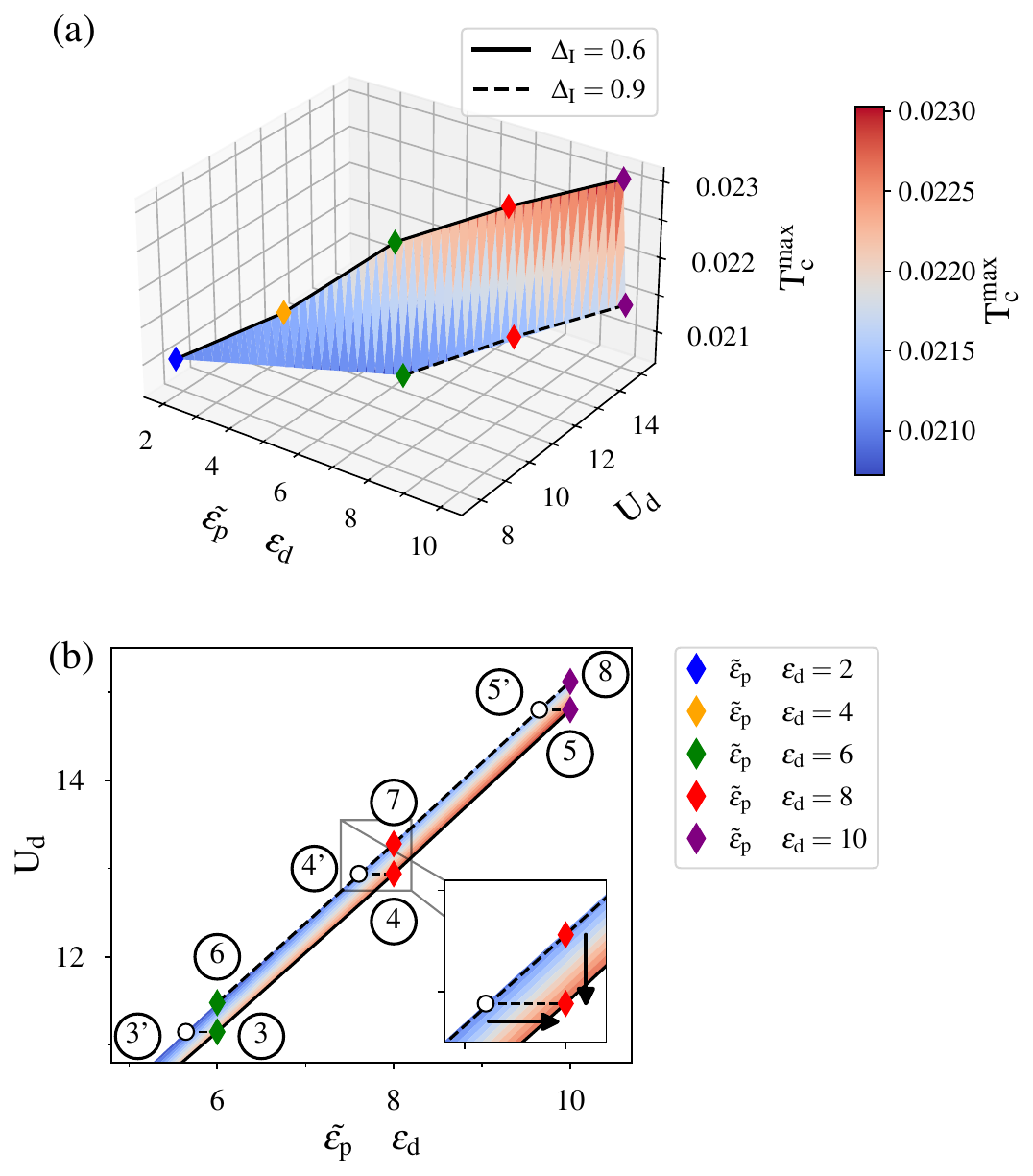}
}
\caption{Same as Fig.~\ref{fig:TcmaxVSDeltaU}, but vs bare Cu-O energy distance $\tilde{\epsilon}_p -\epsilon_d$ rather than $\Delta$. Panel (b) has equal aspect ratio for the $x$ and $y$ axes and covers the data in the range $\tilde{\epsilon}_p -\epsilon_d \in [6, 10]$ only, for a better visualization. For a fixed $\tilde{\epsilon}_p -\epsilon_d$, $T_c^{\rm max}$ increases with decreasing $U_d$ (see e.g. vertical arrow at point labeled $4$). For a fixed $U_d$, $T_c^{\rm max}$ increases with increasing $\tilde{\epsilon}_p -\epsilon_d$ (see e.g. horizontal arrow at point labeled $4$). The percentage changes in $U_d$ and in $\tilde{\epsilon}_p -\epsilon_d$ at data labeled $3$, $4$, and $5$ are computed in Table~\ref{tab:changes}. Diamonds indicate computed data and open circles indicate interpolated data. Inset of panel (b) shows an enlarged view on the contourmap close to the data labeled $4$.
}
\label{fig:TcmaxVSepU}
\end{figure}
\begin{table}
\caption{Percentage change of each model parameter that is necessary to yield a 1\% relative increase in $T_c^{\rm max}$ along a specified path.}
\begin{ruledtabular}
\begin{tabular}{lll}
Path & \% change in $U_d$  & \% change in $\tilde{\epsilon}_p -\epsilon_d$ \\
\hline
\Circled{8} to \Circled{5} & -0.29  &  0  \\
\Circled{5'} to \Circled{5} & 0  & 0.45  \\
\hline
\Circled{7} to \Circled{4} &  -0.33 & 0 \\
\Circled{4'} to \Circled{4} &  0 & 0.59  \\
\hline
\Circled{6} to \Circled{3} &  -0.36 & 0 \\
\Circled{3'} to \Circled{3} & 0  & 0.72 \\
\end{tabular}
\end{ruledtabular}
\label{tab:changes}
\end{table}

Since $\Delta = \epsilon_d + U_d - \tilde{\epsilon}_p$ depends on $U_d$, it is useful to analyze $T_c^{\rm max}$ vs the two {\it independent} model parameters $U_d$ and $\tilde{\epsilon}_p -\epsilon_d$. This is shown in Fig.~\ref{fig:TcmaxVSepU}, with the same color code of Fig.~\ref{fig:TcmaxVSDeltaU}. For a better visualization,  the $x$ axis of Fig.~\ref{fig:TcmaxVSepU}(b) covers the datapoints (diamonds) in the range $[6, 10]$ only. Note also that the $x$ and $y$ axes have equal aspect ratio. For a fixed charge gap size $\Delta_{\rm I}$, $T_c^{\rm max}$ increases with increasing $\tilde{\epsilon}_p -\epsilon_d$). For a fixed $\tilde{\epsilon}_p -\epsilon_d$, $T_c^{\rm max}$ increases with decreasing $U_d$. 

Having identified the interaction strength $U_d$ and the bare Cu-O energy distance $\tilde{\epsilon}_p -\epsilon_d$ as two independent model parameters controlling $T_c^{\rm max}$, we can quantify the rate of change of $T_c^{\rm max}$ vs $U_d$ and $\tilde{\epsilon}_p -\epsilon_d$. 
To do that, we follow Ref.~\cite{Nicolas:PNAS2021} and compute the percentage change of each model parameter that is necessary to yield a $1\%$ relative increase in $T_c^{\rm max}$, along a specified path. This is illustrated in Table~\ref{tab:changes} for three reference datapoints, marked by the circled numbers 3, 4, and 5. The inset of Fig.~\ref{fig:TcmaxVSepU}(b) shows an example of the two paths toward point $4$ (black arrows). 
To obtain the same $1\%$ relative increase in $T_c^{\rm max}$, it is needed to change $U_d$ less than it would need to change $\tilde{\epsilon}_p -\epsilon_d$, by a factor $|-0.29/0.45| \approx 0.64$, $|-0.33/0.59| \approx 0.56$, and $|-0.36/0.72| \approx 0.5$. First, this implies that in all cases, the relative change in $U_d$ is more effective than a relative change in  $\tilde{\epsilon}_p -\epsilon_d$ to yield a relative increase of $1\%$ in $T_c^{\rm max}$. Second, at least for the range of parameters studied in Table~\ref{tab:changes}, the ratio of relative changes in $U_d$ and $\tilde{\epsilon}_p -\epsilon_d$  decreases with decreasing $\tilde{\epsilon}_p -\epsilon_d$. Physically, this implies that the efficacy of changing $U_d$ on $T_c^{\rm max}$ decreases in going from point $3$ to $5$, as expected since the Emery model moves deeper into the charge-transfer regime.

To summarize this section, we studied the $T_c^{\rm max}$ dependence on model parameters $U_d$ and $\Delta$ (or $\tilde{\epsilon}_p -\epsilon_d$). We find that $U_d$ is the dominant variable in changing $T_c^{\rm max}$. This and the companion work~\cite{Eleanor_short} establishes a pathway for optimizing $T_c^{\rm max}$ through doping a charge-transfer insulator which is located deep in the charge-transfer regime and close to the metal-insulator boundary. These findings provide testable predictions for recent proposals to implement the Emery model with ultracold atoms in optical lattices~\cite{Lange2026, McCabe2026}. The realization of low temperatures in cold atom simulators - a necessary step for detecting superconducting correlations - is achievable~\cite{Xu:Nature2025}.

\subsection{$T_c^{\rm max}$ vs physical quantities $\Delta_{\rm I}$ and $2p_p$}

Next, we analyze $T_c^{\rm max}$ versus two physical quantities: charge gap size $\Delta_{\rm I}$  and O hole content $2p_p$.

\subsubsection{$T_c^{\rm max}$ vs charge gap size $\Delta_{\rm I}$ for given values of bare Cu-O energy distance}

\begin{figure}[t!]
\centering{
\includegraphics[width=1.0\linewidth]{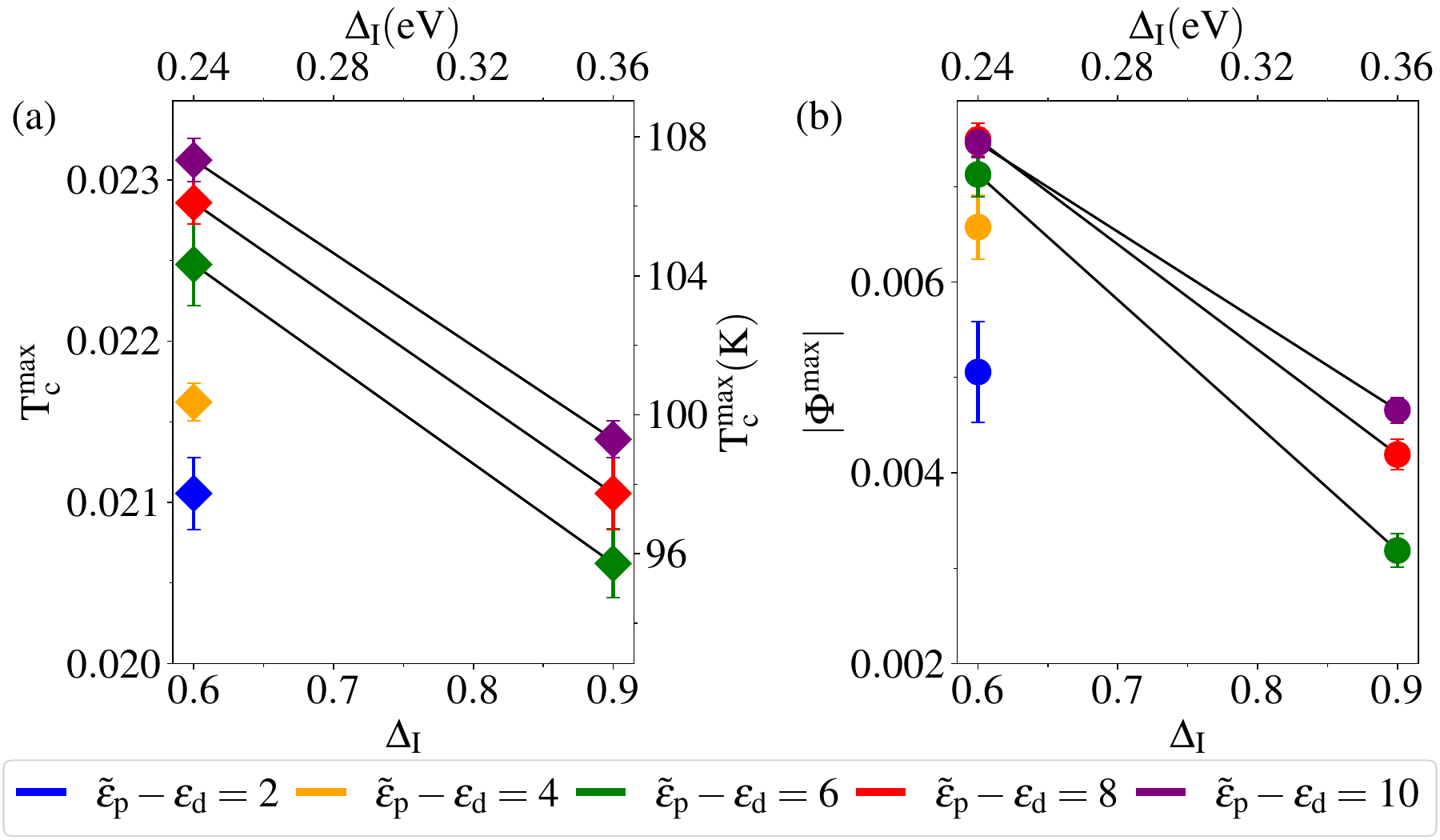}
}
\caption{(a) Maximum superconducting critical temperature $T_c^{\rm max}$ vs charge gap size $\Delta_{\rm I}$ for different values of bare Cu-O energy distance $\tilde{\epsilon}_p -\epsilon_d$. (b) Same as (a), but with maximum superconducting order parameter $\Phi^{\rm max}$ at $T=1/50$ instead of $T_c^{\rm max}$ on the $y$ axis. Data are converted into physical units by setting $t_{pp}=0.4$~eV. Each data point shows the coordinates $(\Delta_{\rm I}, T_c^{\rm max})$ [$(\Delta_{\rm I}, \Phi^{\rm max})$] found in each panel of Fig.~\ref{fig:Tcvsdoping} [Fig.~\ref{fig:SCopvsdoping}]. 
}
\label{fig:TcmaxvsDeltaI}
\end{figure}

Fig.~\ref{fig:TcmaxvsDeltaI}(a) shows $T_c^{\rm max}$ as a function of the charge gap size $\Delta_{\rm I}$ in the parent $\delta=0$ state, for given values of the bare Cu-O energy distance $\tilde{\epsilon}_p -\epsilon_d$. The color code notation is the same as in Fig.~\ref{fig:TcmaxvsDelta}. For a given $\tilde{\epsilon}_p -\epsilon_d$, $T_c^{\rm max}$ decreases with increasing $\Delta_{\rm I}$. This behavior follows from the fact that the charge gap size is controlled by $U_d$. The rate of decrease in $T_c^{\rm max}(\Delta_{\rm I})$ does not vary significantly with $\tilde{\epsilon}_p -\epsilon_d$, but further data are needed to clarify this issue.

The decrease of $T_c^{\rm max}$ with increasing $U_d$, and hence $\Delta_{\rm I}$, was inferred by the decrease of the maximum superconducting order parameter $\Phi^{\rm max}|_{T=0}$ with $U_d$, and hence $\Delta_{\rm I}$, at fixed $\tilde{\epsilon}_p -\epsilon_d$ in Refs.~\cite{Dash:PRB2019, Nicolas:PNAS2021, St-Cyr:2025}. Fig.~\ref{fig:TcmaxvsDeltaI}(b) shows $\Phi^{\rm max}|_{T=1/50}$ vs $\Delta_{\rm I}$. Both $\Phi^{\rm max}|_{T=1/50}$ and $T_c^{\rm max}$ decrease with increasing $\Delta_{\rm I}$, although their rates of decrease differ. Our work extends the validity of the findings of Refs.~\cite{Dash:PRB2019, Nicolas:PNAS2021, St-Cyr:2025} to a broader range of bare Cu-O energy distance.

\subsubsection{$T_c^{\rm max}$ vs oxygen hole content $2p_p$ for given values of charge gap size $\Delta_{\rm I}$}

\begin{figure}[t!]
\centering{
\includegraphics[width=1.0\linewidth]{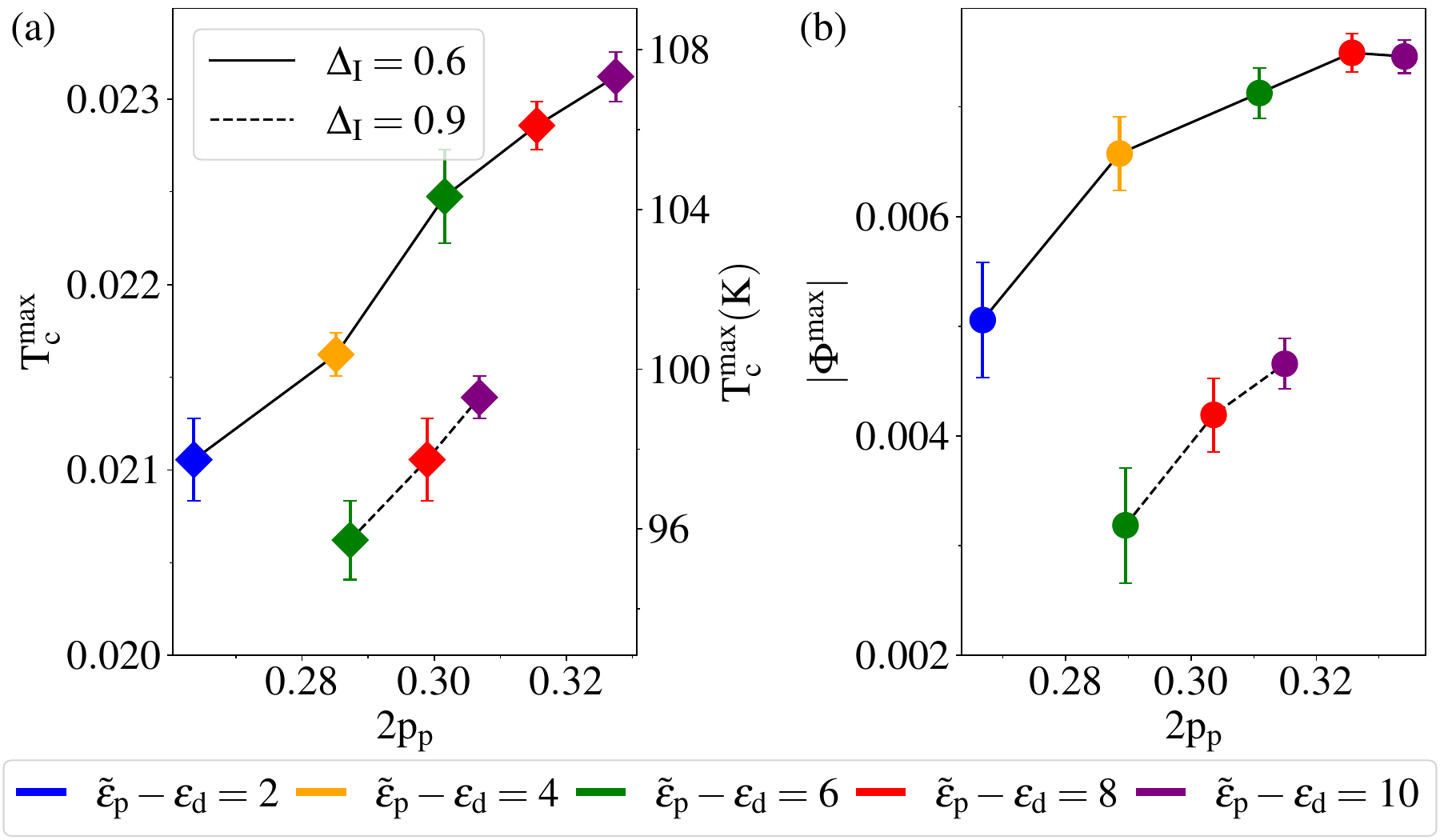}
}
\caption{(a) Maximum superconducting critical temperature $T_c^{\rm max}$ vs O hole content $2p_p$ for two values of charge gap size $\Delta_{\rm I}$. Each data point shows the coordinates $((2p_p)^{\rm opt}, T_c^{\rm max})$ found in Fig.~\ref{fig:TcvsOHC}, where $(2p_p)^{\rm opt}$ is the O hole content optimizing $T_c^{\rm CDMFT}(2p_p)$. Temperature is converted into Kelvin with $t_{pp}=0.4$~eV. (b) Same as (a), but each data point shows the coordinates $((2p_p)^{{\rm opt}, \Phi}, \Phi^{\rm max})$ found in Fig.~\ref{fig:TcvsOHC}, where $(2p_p)^{{\rm opt}, \Phi}$ is the O hole content optimizing $\Phi |_{T=1/50}(2p_p)$. 
}
\label{fig:TcmaxvsOHC}
\end{figure}

Figure~\ref{fig:TcmaxvsOHC}(a) shows $T_c^{\rm max}$ as a function of O hole content $2p_p$ for two given values of the charge gap size $\Delta_{\rm I}$ in the parent state, $\Delta_{\rm I}=0.6$ (solid line) and $\Delta_{\rm I}=0.9$ (dashed line). Each diamond symbol corresponds to the coordinates $((2p_p)^{\rm opt}, T_c^{\rm max})$ found in Fig.~\ref{fig:TcvsOHC}, with the same color code, where $(2p_p)^{\rm opt}$ is the O hole content optimizing $T_c^{\rm CDMFT}(2p_p)$ (see also Appendix~\ref{sec:OHC_appendix}). 

For a given $\Delta_{\rm I}$, $T_c^{\rm max}$ increases with increasing O hole content. Physically this shows that increasing $T_c^{\rm max}$ correlates with the transfer of electrons from O to Cu orbitals, i.e. with increasing the $d$-$p$ mixing of the orbitals. As shown in Fig.~\ref{fig:TcmaxvsOHC}(b), the behavior of $\Phi^{\rm max}|_{T=1/50} (2p_p)$ indicates similar trends as $T_c^{\rm max}(2p_p)$, with different rates of increase. 

The relationship between $T_c^{\rm max}$ and $2p_p$ is a key finding of the companion paper~\cite{Eleanor_short}, where it is shown for $\Delta_{\rm I}=0.6$ only. Here we confirm that this trend is not limited to small values of $\Delta_{\rm I}$, but also holds for larger values of $\Delta_{\rm I}$, as shown by the data at $\Delta_{\rm I}=0.9$. 
Most importantly, the comparison between the results for $\Delta_{\rm I}=0.6$ and $\Delta_{\rm I}=0.9$ allows us to obtain the main physical finding of this work: oxygen hole content $2p_p$ and charge gap size $\Delta_{\rm I}$ are two mechanisms controlling $T_c^{\rm max}$. If $2p_p$ were the only mechanism, the line for $\Delta_{\rm I}=0.6$ and $\Delta_{\rm I}=0.9$ would overlap. Conversely, if $\Delta_{\rm I}$ were the only mechanism, then there would be two parallel curves with zero slope. 
Fig.~\ref{fig:TcmaxvsOHC}(a) shows that the same value of $2p_p$ can yield different values of $T_c^{\rm max}$, and conversely the same $T_c^{\rm max}$ can be obtained with different values of $2p_p$. Hence $\Delta_{\rm I}$ and $2p_p$ are both mechanisms controlling $T_c^{\rm max}$.

\subsubsection{Discussion}

\begin{figure}[t!]
\centering{
\includegraphics[width=1.0\linewidth]{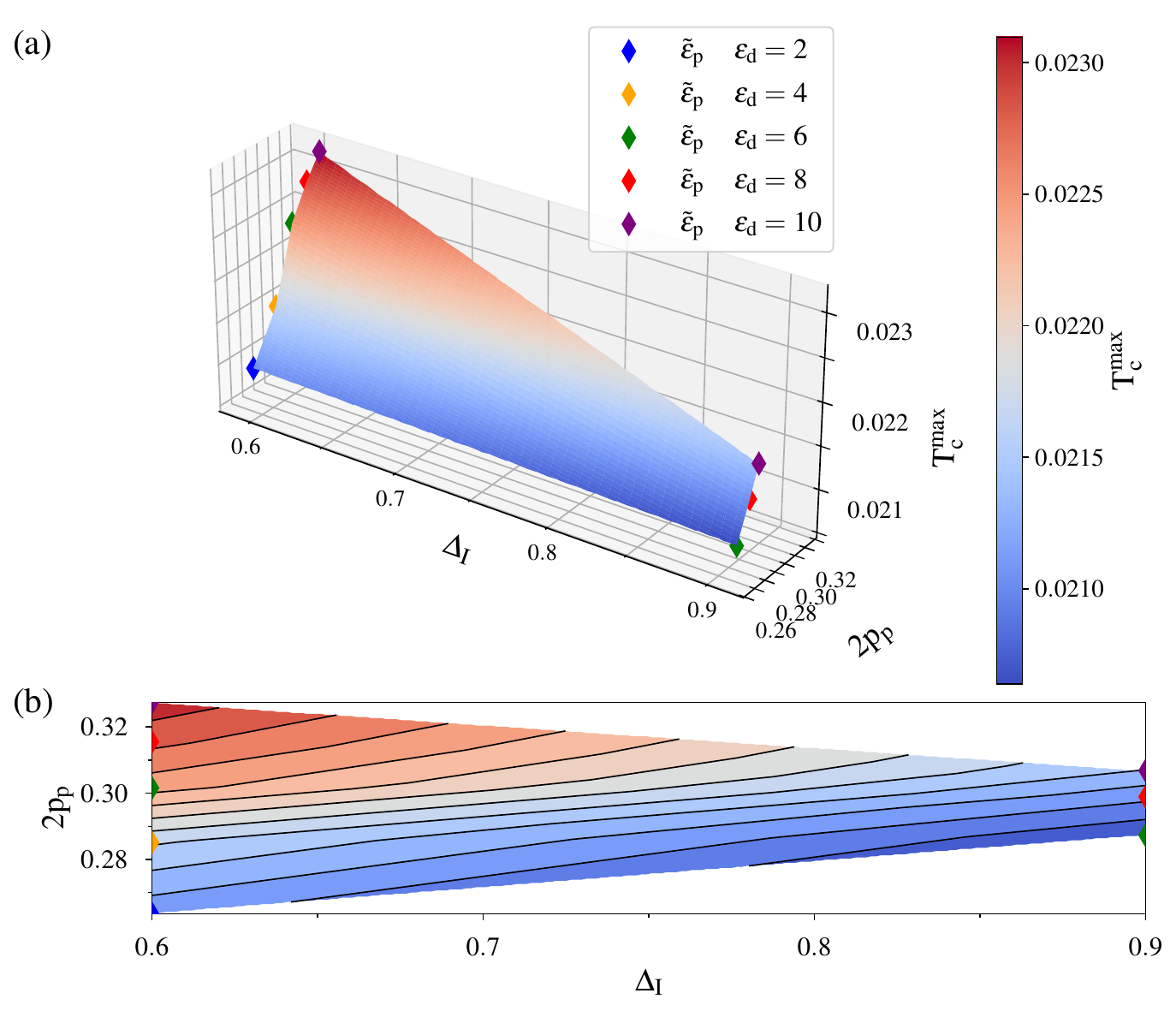}
}
\caption{Same as Fig.~\ref{fig:TcmaxVSepU}, but vs charge gap size $\Delta_{\rm I}$ and oxygen hole content $2p_p$. Both panels have equal aspect ratios for the $x$ and $y$ axes. 
}
\label{fig:TcmaxSummary}
\end{figure}

Oxygen hole content and charge gap size are two quantities governing $T_c^{\rm max}$. We can now discuss their interplay, which forms the key result of this work. Fig.~\ref{fig:TcmaxSummary}(a) shows $T_c^{\rm max}$ as a function of $\Delta_{\rm I}$ and $2p_p$. We obtain the surface $T_c^{\rm max}(\Delta_{\rm I}, 2p_p)$ by interpolating our data (diamonds) on a uniform grid. The color gradient indicates the magnitude of $T_c^{\rm max}$. Note that the $x$ and $y$ axes have equal aspect ratio: we vary $\Delta_{\rm I}$ in the range $[0.6, 0.9]$, corresponding to an absolute change of $0.3$ and $2p_p$ varies approximately in the range $[0.26,0.33]$, corresponding to an absolute change of $0.07$. The resulting contourmap of $T_c^{\rm max}$ is shown in Fig.~\ref{fig:TcmaxSummary}(b). 

In the range of parameters investigated in this study, the largest $T_c^{\rm max}$ occurs under two conditions: for a small charge gap size $\Delta_{\rm I}$ and for a large O hole content $2p_p$. $T_c^{\rm max}(\Delta_{\rm I}, 2p_p)$ monotonically increases with decreasing $\Delta_{\rm I}$ and increasing $2p_p$. In terms of the Zaanen-Sawatzky-Allen diagram, the region close to the metal-insulator boundary and deep in the charge-transfer regime is the region with larger O hole content, where it is most energetically favorable to transfer electrons from O to Cu orbitals. From Fig.~\ref{fig:TcmaxSummary}(b), it is also clear that if $\Delta_{\rm I}$ were the only mechanism governing $T_c^{\rm max}$, then the contour lines of $T_c^{\rm max}$ would be vertical, and conversely if $2p_p$ were the only mechanism governing $T_c^{\rm max}$, then the contour lines of $T_c^{\rm max}$ would be horizontal. There is an overall increase of $T_c^{\rm max}$ with increasing $2p_p$ and with decreasing $\Delta_{\rm I}$.

Our results are compatible with Ref.~\cite{Nicolas:PNAS2021}, which analyzes $\Phi^{\rm max}$ versus oxygen hole content at a constant value of bare Cu-O energy distance or at a constant value of $U_d$ (see also Appendix~\ref{sec:comparison}). These two paths can be seen as given directions across the surface in the $2p_p - \Delta_{\rm I}$ parameter space, and as such the surface in Fig.~\ref{fig:TcmaxSummary}(a) generalises their results. More data are needed to assess if these trends persist in a broader region of $2p_p$ and $\Delta_{\rm I}$. 

These findings give testable predictions for cold atom simulators of the Emery model~\cite{Lange2026, McCabe2026}. Furthermore, these results offer a theory framework for interpreting two key experimental trends in hole-doped cuprates: the anticorrelation of $T_c^{\rm max}$ with charge gap size~\cite{Ruan:SciBull2016, Davis:PNAS2022, Wang:Science2023} and the positive correlation of $T_c^{\rm max}$ with O hole content~\cite{Rybicki:NatComm2016, Jurkutat:PNAS2023}. Together, they indicate that optimizing $T_c^{\rm max}$ requires a narrow charge gap and a large O hole content, as shown in Fig.~\ref{fig:TcmaxSummary}. In addition, our finding of $2p_p$ as the main driver for the change in $T_c^{\rm max}$ indicates that the experimental correlation of $T_c^{\rm max}$ with $2p_p$~\cite{Rybicki:NatComm2016, Jurkutat:PNAS2023} is robust against small variations of $\Delta_{\rm I}$.

\subsection{Strengths and limitations}
At theory level, there are four main limitations in our study of the $T_c^{\rm max}$ trends, which present avenues for future research. First, we calculate the superconducting critical temperature at CDMFT level. $T_c^{\rm CDMFT}$ is the temperature below which the superconducting order parameter $\Phi$ is nonzero. Whilst this calculation advances previous work relying on $\Phi^{\rm max}$ as a proxy for $T_c^{\rm max}$, our method neglects Kosterlitz-Thouless phase fluctuations. These fluctuations can reduce $T_c^{\rm max}$ and can vary its corresponding optimal doping and O hole content, thereby altering the $T_c^{\rm max}$ trends. 

Second, competition with other states, such as antiferromagnetism, is not addressed in this study. This can modify $T_c^{\rm max}$ and can alter its corresponding optimal doping and O hole content, and hence change the $T_c^{\rm max}$ trends. 

Third, here we navigate the large dimensional parameter space of the Emery model by fixing the hopping amplitudes $t_{pp}^\prime$ and $t_{pd}$. As mentioned in Sec.~\ref{sec:Introduction}, this approach results in a two-dimensional optimization problem, i.e. we optimize $T_c^{\rm max}$ as a function of $U_d$ and $\Delta$. In general, it is expected that $T_c^{\rm max}$ depends on $t_{pp}^\prime$ and $t_{pd}$ as well~\cite{Pavarini:PRL2001, Weber2011, Dash:PRB2019, Nicolas:PNAS2021, Vucicevic:PRB2024, St-Cyr:2025}, as well as longer-range hopping amplitudes~\cite{Jacob:2026}.  Therefore, our study should be considered as an entry point for further work optimizing $T_c^{\rm max}$ in the large dimensional parameter space of the Emery model. Note that this limitation does not alter the key conclusion that $T_c^{\rm max}$ is maximized deep in the charge-transfer regime and close to the metal to insulator boundary. Furthermore, the Zaanen-Sawatzky-Allen diagram remains a useful framework for exploring the full parameter space of the Emery model. 

Fourth, we use a minimal cluster that describes $d$-wave superconductivity, consisting of 4 correlated Cu sites. Further work is needed to characterize the dependence of $T_c$ on the cluster size. Yet, the results of our work answer questions of principle (i.e. the main $T_c$ trends, not the precise value of $T_c$) that should be robust against the cluster size when the value of $U_d$ is large. This is because the two most important energy scales, the interaction strength $U_d$ and the superexchange are present in this minimal cluster. One of the difficulties of studying larger cluster sizes is the fermionic sign problem, which worsens exponentially with increasing the cluster size. At the temperatures where superconductivity occurs, the average  Monte Carlo sign is already of order $0.1$ for our minimal cluster (see Appendix~\ref{sec:sign}), therefore larger cluster sizes may yield a severe sign problem.

When comparing our $T_c^{\rm max}$ trends with experiments, a few considerations are in order. First, our results can be tested with recently proposed realizations of the Emery model with ultracold atoms in optical lattices~\cite{Lange2026, McCabe2026}. This is because of the high tunability of cold atoms simulators. Second, for hole-doped cuprate materials, the goal of model calculations is to reveal and understand the dominant mechanisms behind complex behavior with minimal assumptions. Here we showed that the Emery model captures at a qualitative level the increase of $T_c^{\rm max}$ with increasing O hole content and with decreasing the charge gap size, and provides a framework for interpreting these trends. 

Quantitatively, these  $T_c^{\rm max}$ trends show a few limitations. 
First, in our work we fixed the hopping amplitudes $t_{pp}^\prime$ and $t_{pd}$, whereas those values change in different materials~\cite{Weber2011}. 
Second, in our work oxygen hole content varies approximately in the range $2p_p \in [0.26,0.33]$, whereas in experiments it varies over a broader range~\cite{Rybicki:NatComm2016, Jurkutat:PNAS2023}, from nearly $0.25$ of La$_{2-x}$Sr$_x$CuO$_4$ (LSCO) to nearly $0.47$ of YBa$_2$Cu$_3$O$_{6+y}$ (YBCO), and to nearly $0.61$ of a Tl-based material. Hence, our set of parameters appears to quantitatively capture the values of LSCO only. 
Third, the layer differentiation of real materials is taken into account in our model only indirectly, through the charge gap size $\Delta_{\rm I}$. Ab initio investigations are required to address the layer effect on $T_c$~\cite{Cui:NatComm2025, BacqLabreuil:PRX2025} and of the role of the apical oxygens~\cite{Weber2011, Vadnais:2026}.  
Fourth, the slope of the $p_d$ versus $2p_p$ curves (see Fig.~\ref{fig:CHOvsOHC}) is opposite to the experimental trend~\cite{Rybicki:NatComm2016, Jurkutat:PNAS2023}, as already noticed in the normal state in Refs.~\cite{GiovanniPRB2025, GemmaPRB2025}. 
Fifth, setting $t_{pp}=0.4$~eV (secondary axis in Figs.~\ref{fig:TcmaxvsDeltaI} and \ref{fig:TcmaxvsOHC}) yields a $T_c^{\rm max}$ value approximately two times higher than the experimental data for LSCO and a charge gap size $\Delta_{\rm I}$ about three times smaller than typical experimental values. The optimal doping is also underestimated by about a factor of three. Whilst our results indicate that moving away from the metal-insulator boundary will improve the comparison with experiments, the sign problem prevents us from reaching $T_c^{\rm max}$ in this region.

\section{Summary}
\label{sec:summary}
We studied the maximum superconducting critical temperature $T_c^{\rm max}$ obtained upon hole doping a charge-transfer insulator in the Emery model. 

First, we showed that the Zaanen-Sawatzky-Allen diagram, mapping the normal state in the $U_d - \Delta$ plane, enables an intuitive understanding of the relationship between model parameters and physical observables. The location of the charge-transfer insulator on the Zaanen-Sawatzky-Allen diagram is an indicator of the maximum superconducting critical temperature $T_c^{\rm max}$. 

Second, by examining $T_c^{\rm max}$ as a function of $U_d$ and $\Delta$ (or $\tilde{\epsilon}_p -\epsilon_d$), we brought together the results of the companion paper~\cite{Eleanor_short} with insights from prior research to find that $T_c^{\rm max}$ is optimized close to the charge-transfer insulator to metal boundary and, unexpectedly, deep into the charge-transfer regime. We quantify the relative importance of these models parameters in varying  $T_c^{\rm max}$, showing that the main direction of change is along $U_d$, and not $\tilde{\epsilon}_p -\epsilon_d$. 

Third, we rationalized the above finding by connecting the model parameters $U_d$ and $\Delta$ to physical observables, specifically the charge gap size $\Delta_{\rm I}$ and the O hole content $2p_p$. We found that charge gap size $\Delta_{\rm I}$ and O hole content $2p_p$ are two mechanisms governing $T_c^{\rm max}$. By studying $T_c^{\rm max}$ as a function of $\Delta_{\rm I}$ and $2p_p$ (Fig.~\ref{fig:TcmaxSummary}), (i) we find that the locus of optimum $T_c^{\rm max}$ occurs for large O hole content and narrow charge gap size and (ii) we uncover the dependencies between $\Delta_{\rm I}$ and $2p_p$, with $T_c^{\rm max}$ monotonically increasing with increasing $2p_p$ and decreasing $\Delta_{\rm I}$. 

Our results have implications for the theoretical design of strongly correlated electron materials and provide predictions for finding and optimizing $T_c$ with proposed realizations of the Emery model with ultracold atoms in optical lattices~\cite{Lange2026, McCabe2026}. Furthermore, our results provide a framework for rationalizing two key experimental $T_c$ trends, and their resulting interplay, in hole-doped cuprates: the anticorrelation of $T_c^{\rm max}$ with the size of the parent insulating charge gap~\cite{Ruan:SciBull2016, Davis:PNAS2022, Wang:Science2023} and the positive correlation of $T_c^{\rm max}$ with O hole content~\cite{Rybicki:NatComm2016, Jurkutat:PNAS2023}.

\begin{acknowledgments}
We thank P. S\'emon for sharing his continuous-time quantum Monte Carlo code. We thank G. Reaney for discussions and work on a related project~\cite{GemmaPRB2025}. This work has been partially supported by the Canada First Research Excellence Fund. A.-M. S. T. benefits from \href{https://doi.org/10.69777/309032}{RQMP membership}. Simulations were performed on computers provided by the Canada Foundation for Innovation, Calcul Qu\'ebec, and Digital Research Alliance of Canada.
\end{acknowledgments}

\appendix

\section{$T_c^{\rm max}$ vs oxygen hole content for given values of the charge gap size $\Delta_{\rm I}$}
\label{sec:OHC_appendix}

\begin{figure}[ht!]
\centering{
\includegraphics[width=1.0\linewidth]{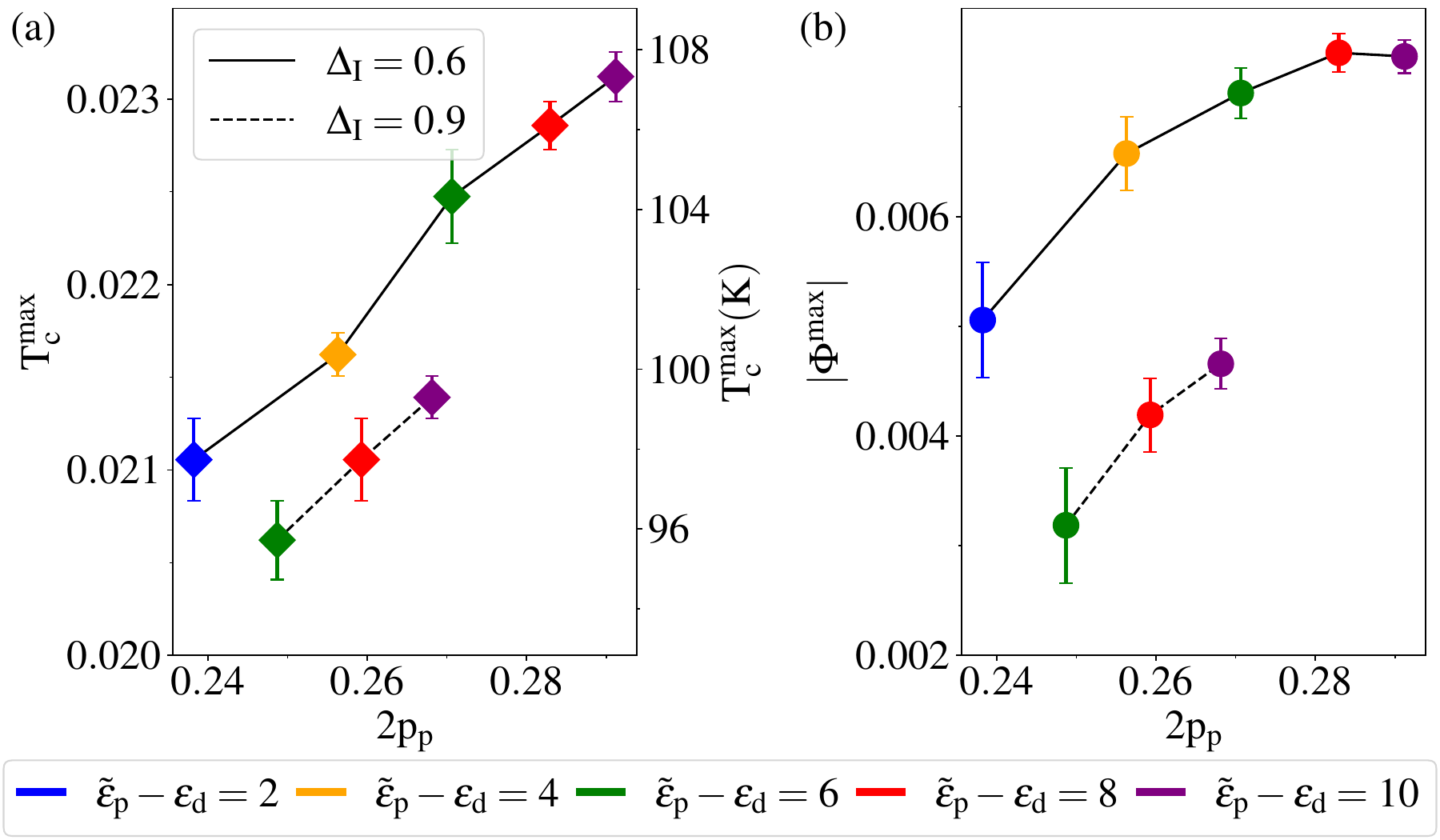}
}
\caption{Same as Fig.~\ref{fig:TcmaxvsOHC}, but here in both panels the $x$ coordinate of each data point shows $(2p_p)^{\delta=0}$, i.e. the O hole content of the parent undoped state at $T=1/50$ found in Fig.~\ref{fig:CHOvsOHC}.
}
\label{fig:TcmaxvsOHCdelta0}
\end{figure}

In Fig.~\ref{fig:TcmaxvsOHC} of the main text, we follow the experimental Refs.~\cite{Rybicki:NatComm2016, Jurkutat:PNAS2023} and show $T_c^{\rm max}$ as a function of O hole content, where we use the {\it optimal} O hole content $(2p_p)^{\rm opt}$, i.e. the O hole content that optimizes $T_c^{\rm CDMFT}$, as the $x$ coordinate of each data point. 

It is interesting to show $T_c^{\rm max}$ as a function of O hole content using instead the O hole content {\it at zero doping} $(2p_p)^{\delta=0}$ as the $x$ coordinate of each data point. This is shown in Fig.~\ref{fig:TcmaxvsOHCdelta0}. Both Fig.~\ref{fig:TcmaxvsOHCdelta0} and Fig.~\ref{fig:TcmaxvsOHC} show similar trends. 

Note that $T_c^{\rm max}$ is a quantity found at finite doping (see e.g. Fig.~\ref{fig:Tcvsdoping}), whereas $(2p_p)^{\delta=0}$ is a quantity at zero doping. This is similar to Fig.~\ref{fig:TcmaxvsDeltaI}, where we plot $T_c^{\rm max}$ against the charge gap size $\Delta_{\rm I}$ in the undoped state. Therefore it is even more remarkable that the parent insulating state provides an indication of the superconducting critical temperature, since both $\Delta_{\rm I}$ and $(2p_p)^{\delta=0}$ are indicative of $T_c^{\rm max}$ trends.

\section{Comparison with previous CDMFT work}
\label{sec:comparison}

\begin{figure}[t!]
\centering{
\includegraphics[width=1.0\linewidth]{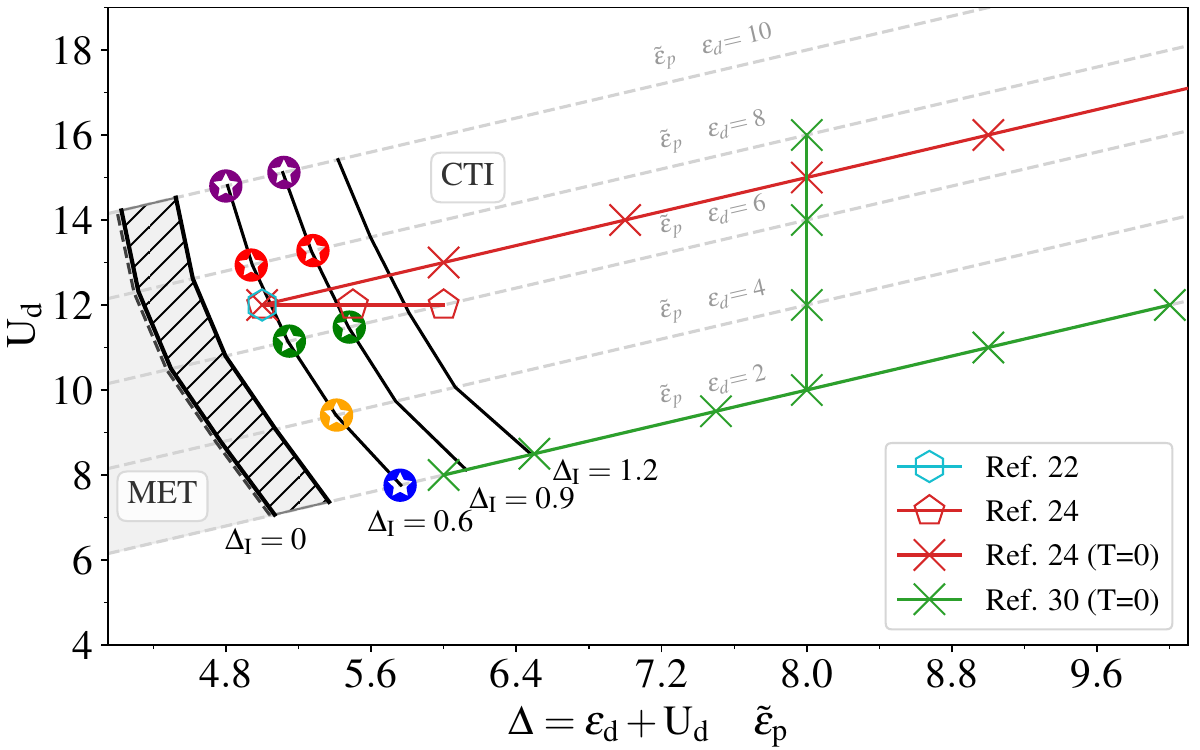}
}
\caption{Same as Fig.~\ref{fig:zsa}(a), with only the data points of the parent state from which the superconducting state has been computed. Circles with star indicate our results. Other symbols show results of prior CDMFT work~\cite{Lorenzo3band, Nicolas:PNAS2021, St-Cyr:2025} (see legend). 
}
\label{fig:cfr}
\end{figure}

We have shown that the Zaanen-Sawatzky-Allen diagram $U_d - \Delta$ is a valuable framework for exploring and rationalizing the $T_c$ trends in the Emery model. Therefore it is useful to translate into the $U_d-\Delta$ diagram of Fig.~\ref{fig:zsa}(a) the data points investigated in previous CDMFT studies on the superconducting trends in the Emery model~\cite{Weber2011, Lorenzo3band, Dash:PRB2019, Nicolas:PNAS2021, St-Cyr:2025}. 

Since a point in the $U_d - \Delta$ diagram marks the parent state from which the superconducting state emerges, Fig.~\ref{fig:cfr} shows the same Zaanen-Sawatzky-Allen diagram of Fig.~\ref{fig:zsa}(a), where for clarity we indicate only the data points of the parent state from which the superconducting state has been computed. Circles with stars indicate the eight data points studied in our work (see e.g. Fig.~\ref{fig:Tcvsdoping}). The blue hexagon shows the data point studied in Ref.~\cite{Lorenzo3band}. Red pentagons show the three data points analyzed in Ref.~\cite{Nicolas:PNAS2021}. Both Ref.~\cite{Lorenzo3band} and Ref.~\cite{Nicolas:PNAS2021} studied the superconducting state at finite temperature and hence computed $T_c^{\rm max}$. 

Results with crosses report $T=0$ calculations, where the maximum superconducting order parameter $\Phi^{\rm max}$ is used as a proxy for $T_c^{\rm max}$. They include data in Ref.~\cite{Nicolas:PNAS2021} (red crosses) and in Ref.~\cite{St-Cyr:2025} (green crosses). Similar data points  at $T=0$ have been analyzed in Ref.~\cite{Dash:PRB2019}. Note that the data of Ref.~\cite{St-Cyr:2025} (green crosses) use different values of the hopping amplitudes with respect to other data points indicated in Fig.~\ref{fig:cfr}. Similarly to the finite temperature results of Ref.~\cite{Nicolas:PNAS2021} (red pentagons), Ref.~\cite{Weber2011} studied at $T=0$ a path of constant $U_d$ for different values of bare Cu-O energy distance. 

The trends of $T_c^{\rm max}$ as a function of charge gap size $\Delta_{\rm I}$ in the parent $\delta=0$ state, for given values of the bare Cu-O energy distance $\tilde{\epsilon}_p -\epsilon_d$ (Fig.~\ref{fig:TcmaxvsDeltaI}) agree and extend the results of Refs.~\cite{Weber2011, Nicolas:PNAS2021} over a broader range of Cu-O energy distance. 

From Fig.~\ref{fig:cfr}, it is clear that before our work, known paths for controlling $T_c$ all involve changing the charge gap size $\Delta_{\rm I}$ in the undoped state. The novel strategy of our work is to study the behavior of $T_c^{\rm max}$ as a function of bare Cu-O energy distance $\tilde{\epsilon}_p -\epsilon_d$ (or bare charge-transfer energy $\Delta$) {\it at constant values of charge gap size} $\Delta_{\rm I}$ in the parent $\delta=0$ state. This allows us to study the effect of both charge gap size and bare Cu-O energy distance, and hence O hole content, on $T_c^{\rm max}$ (see Fig.~\ref{fig:TcmaxSummary}).

\vspace{4mm}
\section{Monte Carlo sign}
\label{sec:sign}

\begin{figure*}[t]
\centering{
\includegraphics[width=1.0\linewidth]{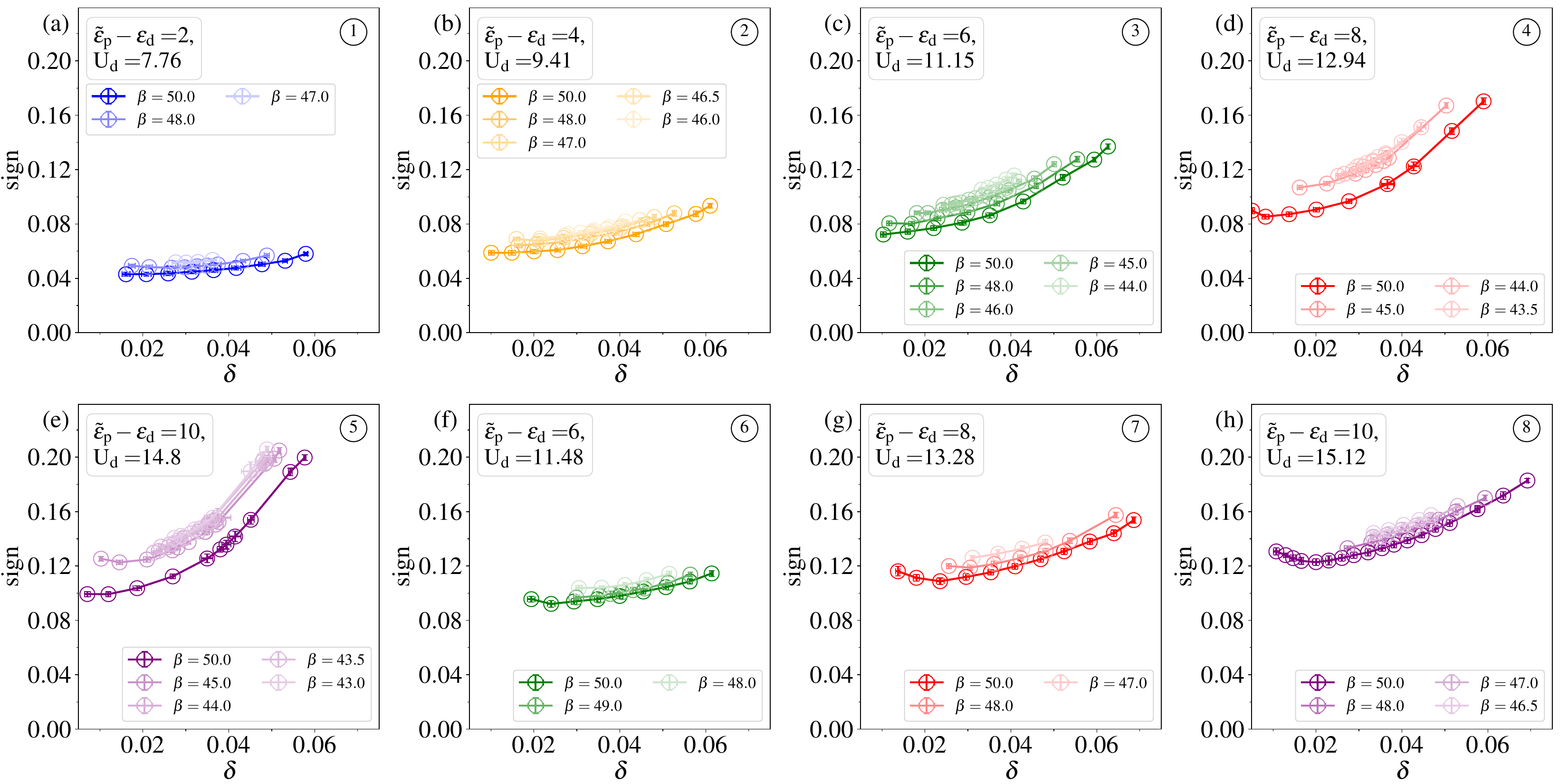}
}
\caption{(a)-(h) Monte Carlo sign versus doping $\delta$ for different values of the inverse temperature $\beta=1/T$ in the superconducting state. Each panel corresponds to the coordinates $(\tilde{\epsilon}_p -\epsilon_d, U_d)$.
}
\label{fig:sign}
\end{figure*}

Figure~\ref{fig:sign} displays the average sign of the continuous-time quantum Monte Carlo simulations versus hole doping and for different temperatures. Data are for the same model parameters of Fig.~\ref{fig:Tcvsdoping}. Note that for a given value of the charge gap size $\Delta_{\rm I}$, the sign overall increases as the system moves deeper into the charge-transfer regime (i.e., from points labeled $1$ to $5$ or $6$ to $8$).

\end{document}